\documentclass[12pt, preprint]{aastex}
\usepackage{amsmath}
\usepackage{graphicx}
\usepackage{subfigure}

\def\cc{cm$^{-3}$}
\def\kms{km s$^{-1}$}

\def\ammonia{NH$_3$}
\def\h2{H$_2$}
\def\n2h{N$_2$H$^+$}
\def\Ms{M$_\odot$}

\def\13co{$^{13}$CO}
\def\c18o{C$^{18}$O}
\def\nh{$n(H_2$)}
\def\lp{\>\> .}
\def\lc{\>\> ,}
\def\cm2{cm$^{-2}$}
\def\nh{n(H$_2$)}

\def\Ms{$M_{\odot}$}
\def\mic{$\mu$m}

\def\h2{H$_2$}

\begin{document}


\title{Massive Quiescent Cores in Orion. I. Temperature Structure}
\author{D. Li\altaffilmark{1,2}, P. F. Goldsmith\altaffilmark{2}, 
	and K. Menten\altaffilmark{3} }

\altaffiltext{1}{Center for Astrophysics, 60 Garden Street, 
Cambridge MA 02138, dli@cfa.harvard.edu }
\altaffiltext{2}{National Astronomy and Ionosphere Center, Department of 
Astronomy, Cornell University,\\ Ithaca NY 14853}
\altaffiltext{3}{Max-Planck-Institut f\"{u}r Radioastronomie, 
	Auf dem H\"{u}gel 69, 53121 Bonn, Germany }

\begin{abstract}
	
We have mapped four massive cores in Orion using the \ammonia\ (J,K) = (1,1) 
and (J,K) = (2,2) inversion transitions, as part of our effort to study the 
pre--protostellar phase of massive star formation. 
These cores were selected to be quiescent, i.e. they contain no apparent 
IR sources and are not associated with any molecular outflows. 
These cores are one order of magnitude more massive than dark cloud cores and 
have about twice the line width. 
This paper focuses on their temperature structure. 
We find a statistically significant correlation between the gas kinetic temperature 
and the gas column density. 
The general trend is for the gas to be colder where the column density is higher,
which we interpret to mean that the interiors of these cores are colder than
the regions surrounding them.
This is in contrast with dark cloud cores, which exhibit relatively flat 
temperature profiles. 
The temperature gradient within the massive quiescent Orion cores is consistent 
with an external radiation source heating the dust, and dust--gas collisions
providing relatively close coupling between dust and gas temperatures.
Thus, we suggest that the initial stage of massive pre--protostellar cloud cores is 
relatively quiescent condensations which are cooler than their surroundings.

\end{abstract}
\setcounter{footnote}{0}

\section{INTRODUCTION}

There appears to be a clear observational dichotomy between low mass star 
formation (LMSF) and high mass star formation (HMSF). 
HMSF only occurs in giant molecular clouds (GMC), while LMSF occurs in a variety 
of environments, including isolated Bok globules \citep{yun90}.
No young star more massive than 2 M$_\odot$ has been found in Taurus, 
while both O stars and low mass stars are abundant in Orion  
\citep*[e.g.\ ][]{muen02}. 
The comparison between the two types of regions shows that regions
with HMSF have a higher efficiency in converting 
interstellar medium (ISM) mass into stellar mass. 
A particularly spectacular star forming theater is the Orion Nebula,
where more than $2\times 10^{3}$ young stars are contained in a radius of about 
0.2 pc centered on the Trapezium O star cluster \citep{jones88, hillenbrand98}.

Star formation is correlated with dense molecular gas
\citep*[e.g.\ ][]{ungerechts87}. 
Particular attention has thus been focused upon
dense structures, so called ``cores'',  within molecular cloud complexes
as they may be sites of future star formation.  
Here, the cores refer to structures usually prominent in high density 
tracers, e.g.~CS and NH$_3$, with mass between that of a single star and of 
a stellar cluster \citep{walmsley95}. 
The cores can be divided into cores in LMSF regions such as Taurus 
\citep*[e.g.\ ][]{onishi98}, and cores in HMSF giant molecular cloud (GMC) 
regions such as Orion (Tatematsu et al.\ 1993). 
The ensemble of cold cores in Orion has twice the mean size (0.3pc), twice the 
line width, and almost an order of magnitude higher mean mass (80 M$_{\odot}$) 
than do the cores in LMSF regions  \citep*[e.g.\ ][]{bm89, cb88}.
	
Can the difference in the parameters of molecular cores help explain the dichotomy 
in star formation?
Theoretically, there is a promising suggestion regarding the initial energy 
balance between gravity and turbulent pressure support.
For LMSF, the picture is relatively clear. 
A core starts in equilibrium, supported by magnetohydrodynamic (MHD) turbulence \citep{myers88}.
The magnetic field slowly decouples from the cloud as a result of ambipolar 
diffusion \citep{shu87, basu95}. 
As the magnetic support weakens, collapse starts from the center (highest
density) and propagates from the inside to the outside of the cloud 
\citep{shu77,larson69,penston69}.
Since the `inside out' collapse propagates roughly at the sound
speed, it is considered a `slow' process, having a time scale
on the order of $10^6$ to $10^{7}$ yr. 
For HMSF, the process needs to be faster, in order to be consistent with 
the observed rate of star formation in a region like Orion. 
The theoretical hypothesis is that the core starts in a {\em super--critical} 
state, in which gravity overwhelms internal support.
Then, the cloud collapses quickly on a free fall time scale
and no ambipolar diffusion is needed. It is worth noting that super--critical
state may also be applicable to low mass cores. 
If a core is embedded in a cloud
complex, the ambient pressure can be larger than a critical value above
which no magnetohydrostatic equilibrium exists. 
For such super--critical
cores, the linewidth is supersonic, the star formation time scale 
is a couple of free--fall times, and no ambipolar diffusion is needed \citep{nakano98}.
This alternative picture (supercritical cores, no ambipolar diffusion) 
of LMSF is more consistent with the empirical evidence concerning
Orion cores than with those of Taurus cores. In Orion, 
LMSF is very common, although HMSF (through the effects of the massive
stars which result) dominates the morphology and conditions of
the region. 
 
Modeling the actual process of HMSF is a very demanding undertaking, due in part
to the difficulties in dealing with non--equilibrium problems and in part 
to the lack of observational constraints. 
There are at least four obstacles to unraveling the complex process of high mass star 
formation. 
Firstly, it is fast. 
A relevant time scale is the stellar Kelvin--Helmholtz (KH) time scale,
\begin{equation}
t_{KH} = \alpha \frac{GM^2}{RL} \lc
\label{kh}
\end{equation}
where $\alpha$ is a factor of the order of unity depending on
the density profile of the object, and $M$, $L$, and $R$ are the stellar mass,
luminosity, and radius, respectively.
From both empirical evidence and theoretical considerations
\citep{clayton83}, there exist scaling laws of the form
 of $L \propto M^{3.5}$  and $R \propto M$,
which simplify Eq.~\eqref{kh} as
\begin{equation}
t_{KH} \approx 19~\text{Myr} ( \frac{M}{M_\odot})^{-2.5} \lp
\label{kh1}
\end{equation}
Increasing the stellar mass from 1 \Ms\ to 10 \Ms\ reduces $t_{KH}$ by a 
factor of 300.
For stars more massive than 10 \Ms, the KH time scales are less than 0.1 Myr.
There is probably no pre--main--sequence phase for massive star formation. 
Secondly, HMSF is dynamic.
Once hydrogen burning starts, the environment of the newly--formed star 
will be severely altered, thus precluding the study of its immediate 
precursor state. 
Thirdly, the molecular condensations possibly leading to HMSF are found
in cloud complexes and are hard to properly identify.
Finally, there are no HMSF regions as close as Taurus, the prototypical region
for studying LMSF star formation.

From an observer's point of view, we want to
tackle the problem from its very beginning, the pre--protostellar phase. 
This study starts with the very basic objective of identifying massive 
cores in a HMSF region.
Fifteen quiescent cores have been chosen from the CS survey of Orion by 
\citet{tatematsu93} based on the following criteria:
	1) We are restricted to a nearby GMC and relatively large cores 
in order to resolve structures with the available beam size of about 50\arcsec.
Fortunately, there is an adequate selection of cores in Orion having diameters 
larger than 0.4 pc or 3'. 
	2) We avoid cores too close to Orion-KL where the energy output of 
young, massive stars has dictated the filamentary morphology of the surrounding 
material \citep{wise98}.
	3) There is some information on the cores in our sample suggesting 
they are relatively isolated and cool (no IR/YSO association), but no 
complete data sets or models have been obtained.

These cores have been mapped in multiple transitions of \c18o and CS,
the J=1--0 transitions of \n2h, the (1,1) and (2,2) inversion transitions
of \ammonia, and in the 350 \mic~ dust continuum. 
The goal of is to retrieve most of the relevant physical 
parameters of these cores, including size,  mass, temperature, and 
density structure. The sources and their derived parameters are listed in
Table 1.
Additional data on these cores including their mass and density, is obtained
from \c18o and CS observations conducted at the Five College Radio
Astronomy Observatory (FCRAO).  These observations will be presented in 
detail in a later paper.
Some information from FCRAO observations is 
included here to aid the analysis of the thermal
structure presented in Sections \ref{therm} and \ref{diss}.
This paper will focus on the temperature structure of these cores
obtained from \ammonia\ observations, while other aspects will be considered
in subsequent work.

\section{Kinetic Temperature and \ammonia \label{kintemp}}
The kinetic temperature is a crucial parameter for defining 
excitation conditions from multi--transition spectral line observations. 
In order to determine the column density accurately from tracers such as
the \c18o\ J=1--0 transition, a good knowledge of the gas temperature 
is also required.
Together with volume density, the kinetic temperature determines the 
level of thermal support inside a cloud. 
Information about the gas temperature is thus a prerequisite for evaluating the 
importance of turbulence. 
Low mass cores have been found to have fairly uniform temperatures 
according to \ammonia\ observations \citep{tafalla02}. 
As modeled, this uniformity of
temperature applies to regions of modest to large extinction,
over which ammonia is observed. As
such, it does not, for example, exclude the possibility of the edges
of dark clouds being heated by the ISRF \citep{snell81,young82}.

The temperature structure of massive cores is not well characterized.
In this study we use the inversion transitions of \ammonia\ to probe 
the Orion core temperatures.
In the remainder of this section we discuss the structure of the ammonia molecule
and how we use it for determining the gas temperature.
In Section \ref{obs} we present the observations, and in Section \ref{struc}
we discuss the temperature structure that we obtain for the HMSF cores.
We analyze the statistical uncertainties associated
with the temperature gradients in Section \ref{stat} . 
In Section \ref{therm} we present a model for the temperature distribution in
these cores based on dust heating by the enhanced radiation field in the
surrounding material.
We discuss our findings and summarize our conclusions in Section \ref{diss}.

\subsection{Characteristics of Ammonia Inversion Transitions \label{nh3hype}}

Ammonia is a symmetric top molecule, which produces
a relatively simple rotational energy level structure. 
The groups of levels having quantum number $K$, 
which describes the angular momentum 
around the symmetry axis through the nitrogen atom, are called
``$K$--ladders''. 
Radiative transitions between $K$--ladders, such as (2,1) to (2,2)
are prohibited. 
Therefore, the relative level population of the (2,2) and (1,1) levels 
is affected only by collisions. 
If the populations of levels in different $K$--ladders can be determined, 
their ratio will be a measure of their excitation, and thus of the temperature of
the collision partners.

The positions of the energy levels are also helpful
in determining the kinetic temperature. 
The metastable levels (lowest level in each $K$--ladder) (1,1)
and (2,2) are at energies equivalent to 23 K and 64 K, respectively, 
above the ground state.
At temperatures characteristic of GMCs, these levels will be significantly 
populated.
Higher J levels, on the other hand, will
be much less populated. 
The population of the (2,1) level, for example, will only be 7.6\% of that 
of the (1,1) level in a cloud thermalized at 20 K. 
Thus, the metastable levels have most of the \ammonia\ molecules in each 
$K$--ladder at moderate temperatures.

Each metastable level for K $\neq$ 0 is split into a pair of so--called
inversion levels.
The frequencies of inversion transitions of the (1,1)
and (2,2) levels are very close, and their proximity 
proves to be a major convenience, since they can be observed simultaneously 
and their relative calibration is largely independent of antenna efficiency
and pointing.
	
The inversion transitions are further split into hyperfine
components due to the coupling between the nuclear spins
of N and H and molecular rotation. As described by
Rydbeck et al.\ (1977), there are 18 components in the
(1,1) inversion transition. 
Usually the individual hyperfine transitions are not resolved and 
only five groups can be identified in a spectrum of an
interstellar cloud. 
The relative strength of the main ($m$) component (the central group) 
and the satellite components ($s$) is determined theoretically under the 
assumption that they have the same level of excitation.
For the (1,1) transition, the optical depths of the five groups conform to
the following relations
\begin{equation}
\begin{split}
\tau^m(1,1)= 3.60 \tau^s(1,1)~\text{(the two inner satellite components)} \lc\\
\tau^m(1,1)= 4.50 \tau^s(1,1)~\text{(the two outer satellite components)} \lc\\
\tau^m(1,1)= 0.50 \tau(1,1) \lc
\end{split}
\label{tau11}
\end{equation}
where $\tau$, $\tau^m$, and $\tau^s$ are the optical depths of
the total, main line, and the individual satellite lines, respectively.
For the (2,2) transition, since only the main line is typically visible, 
the relation of practical use is 
\begin{equation}
\label{tau22}
\tau(2,2)=1.26\tau^m(2,2) \lp
\end{equation}

\subsection{Derivation of The Kinetic Temperature and \ammonia\ Column Density}

For a cloud with temperature around 20 K and about
1 \kms\ doppler broadened line width, we usually can detect 
the hyperfine components of the (1,1) transition in five groups but 
only the main component of the (2,2) transition.

A standard procedure to derive the kinetic temperature through such observations
is discussed by Ho \& Townes (1983). In this procedure, the main and the satellite
antenna temperatures of the (1,1) line and the  main antenna temperature of the (2,2) 
are obtained through fitting the spectra. These three quantities are then combined
with equations~\eqref{tau11} and \eqref{tau22} to derive the relative populations
of the (2,2) and (1,1) levels, defined by the rotation temperature, denoted $T_R$.
$T_R$ can be converted to $T_k$
through excitation calculations, most conveniently empploying the three level approximation
of \citet{walmsley83}.

In our derivation, we follow the same line of argument discussed above, but 
incorporate two recent developments.
First, the optical depths of the (1,1) and (2,2) transitions are not directly fitted 
in deriving $T_R$. 
We use the fitting scheme built into GILDAS, which chooses a
rather particular set of parameters \citep{bachiller87}.
These parameters can be combined to give $T_R$ under the same assumptions
concerning uniform and equal excitation conditions, with this technique being more
stable at low optical depths. 
The column density of the (1,1) level ($N(1,1)$), also a result of this fitting procedure,
is used to calculate the total column density of \ammonia\ 
assuming that only the four lowest levels are populated, and that their
relative populations are all defined by $T_R$, using \citep{rohlfs96}
\begin{equation}
\label{nnh3}
N(\text{\ammonia}) = N(1,1)[\frac{1}{3}\exp(23.1/T_R)+1+\frac{5}{3}\exp(-41.2/T_R)+\frac{14}{3}\exp(-99.4/T_R)] \lp
\end{equation}
The peak \ammonia\ column densites derived are listed in Table 1. 
Note that the higher metastable levels, such
as (4,4), contain less than $1\%$ of the total population.

Second, Danby et al.\ (1988) have calculated the cross sections for
collisions between \h2 and \ammonia\ in states up to J=5. 
These authors also provide results of excitation analysis
for graphically converting $T_R$ to $T_k$. 
We have fitted a polynomial to their curve and use it instead of the 
analytic approximation of \citet{walmsley83}.

The assumption of equal excitation, which underlies all the fitting procedures discussed above,
will bring uncertainties
into the derivation, especially in a low density environment.
According to \citet{stutzki85}, such effects are small for the derived rotational temperature
under the conditions of the Orion cores (\nh$\sim 5\times10^{4}$ \cc). The
uncertainty in $T_k$ is dominated by the noise in the (2,2) spectra, as discussed in
a later section.

The lower excitation of the satellite lines
causes an underestimate of the column 
density derived through equation~\eqref{nnh3} by about $10\%$. 
The exclusion of non-metastable levels introduces
another $\sim 5\%$ underestimate. 
Based on the \ammonia\ column density, the \c18o\ column density (discussed in a later paper), 
and a constant \c18o\ abundance
[\c18o]/[\h2]=$1.7\times10^{-7}$, we find
the average \ammonia\ fractional abundance in
the Orion cores to be $5\times10^{-9}$.

\section{Observations \label{obs}}

The majority of our ammonia data come from the Effelsberg
100m telescope. 
Calibration of the 100m is carried out by observations of point sources.
The raw data are in system units, say counts, for which two factors are
needed for the conversion to antenna temperature units. 
The first factor is `K/counts', relating system units to antenna temperature.
This is obtained using noise sources which have been
calibrated against thermal loads.
The second factor is `K/Jy', which is given by the effective area 
divided by $2k$, with $k$ being Boltzmann's constant.
This factor is obtained by observing point--source calibrators, including 
3C147 \citep{ott94}. 
The limited data obtained using the NRAO
\footnote{The National Radio Astronomy Observatory is a facility of the
National Science Foundation operated under cooperative agreement
by Associated Universities, Inc.
} 
140 foot telescope was also
converted to antenna temperature using separately--calibrated noise diodes.
The aperture efficiency was determined by observations of standard
radio point sources.

Our Orion sources are not point sources relative to the beam, in which
case  the main beam efficiency $\epsilon_{mb}$ is more 
relevant than the aperture efficiency $\epsilon_A$.
For a known beam shape,  $\epsilon_{mb}$ can be obtained from $\epsilon_A$.
The main beam solid angle, $\Omega_{mb}$, defined as 
the integral over the main beam of
the normalized power pattern, can be straightforwardly determined if a
Gaussian satisfactorily represents the main beam.  
Thus, we find that $\Omega_{mb} = 1.13{\theta_{FWHM}}^2$, where $\theta_{FWHM}$
is the full width to half maximum beam width.
The main beam efficiency is then obtained from
\begin{equation}
\epsilon_{mb} = \frac{\Omega_{mb}}{\Omega_A} = \frac{\Omega_{mb} A_p \epsilon_A }
			{\lambda^2} \lc
\end{equation}
where $\Omega_A$ is the antenna solid angle, and we have used the antenna
theorem to relate it to the effective area $A_e$, which is equal to
the product of the physical area $A_p$ and the aperture efficiency. 

At the 140 foot telescope, we measured the beam efficiency to be about 20$\%$.
There is also evidence of dish deformation as the source elevation varies.
For this reason, the data from the 140 foot are not calibrated for their absolute 
intensity. We used our 140 foot time to search for ammonia peaks in our
Orion sources and obtained several kinetic temperature measurements,
which are relatively independent of absolute calibration.

At Effelsberg, only the inner 75m of the 100m antenna is usable
at 24 GHz, giving a FWHM beam size of 43\arcsec. 
We determine the main beam efficiency to be around 0.60 and
the data has been calibrated accordingly.
However, atmospheric effects are not removed, and there were
some
gain variations (Muders 2001, private communication), so that the uncertainty can
be as large as 50\%.

The autocorrelator AK90, at Effelsberg, is divided into
four quadrants, each with 2048 channels and 20MHz bandwidth.
This bandwidth allows frequency switching even for the \ammonia\
hyperfine lines. 
The (1,1) and (2,2) inversion transitions are observed simultaneously
in two linear polarizations, which are then averaged.
The processed data have a velocity resolution of 0.24 \kms.
With single beam pointings, we constructed beam--sampled 
maps at 0.738\arcmin\ spacing to cover the same region as
the FCRAO observations. 
Maps were obtained of the sources ORI1, ORI2, ORI4 and ORI8 (Table 1).

\section{Temperature and Turbulence Structure\label{struc}}
\subsection{Temperature Gradients}
\begin{figure}[htp]
      	\centering 
   	\subfigure{\includegraphics[width=9cm]{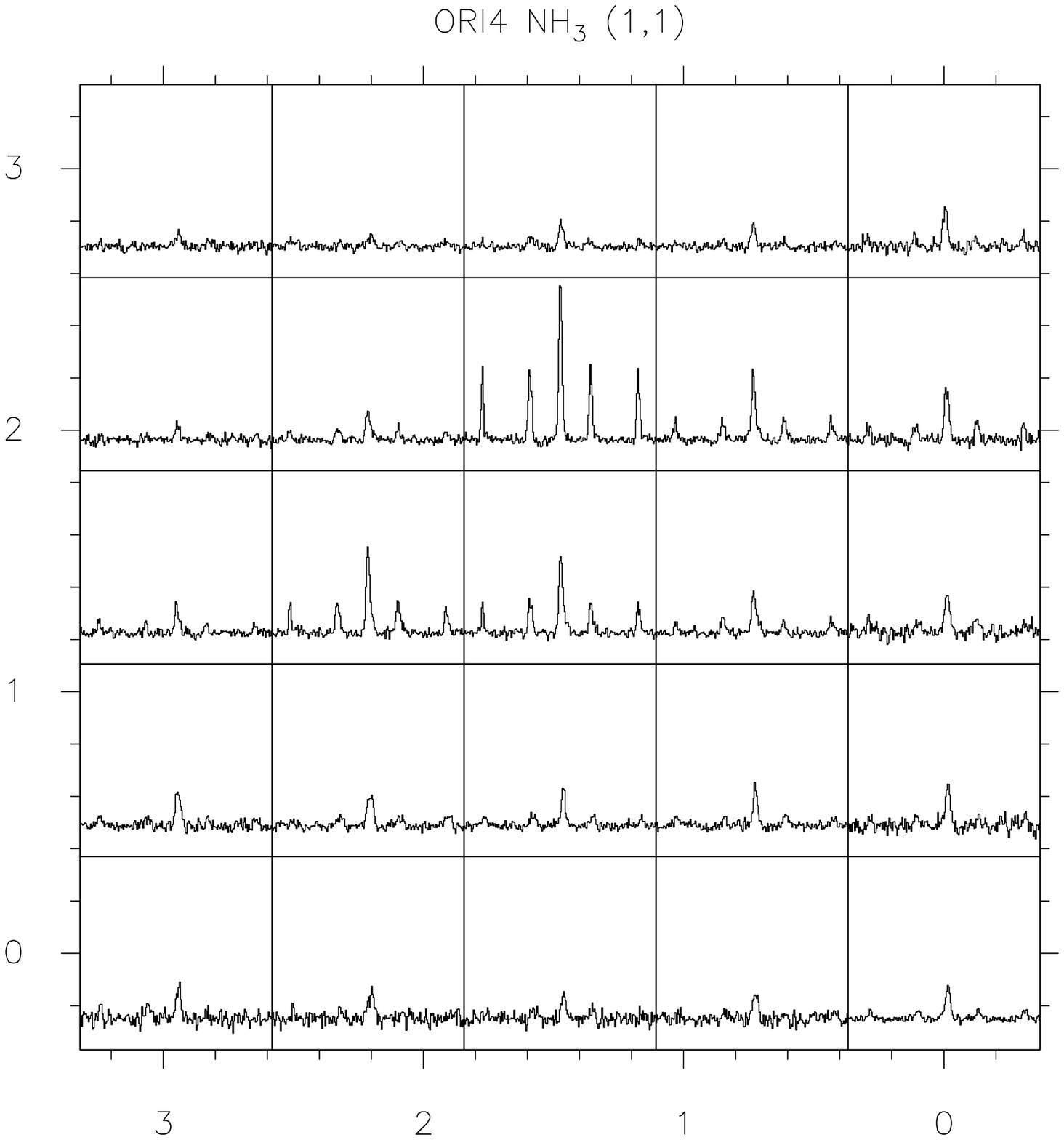}}\\
   	\subfigure{\includegraphics[width=9cm]{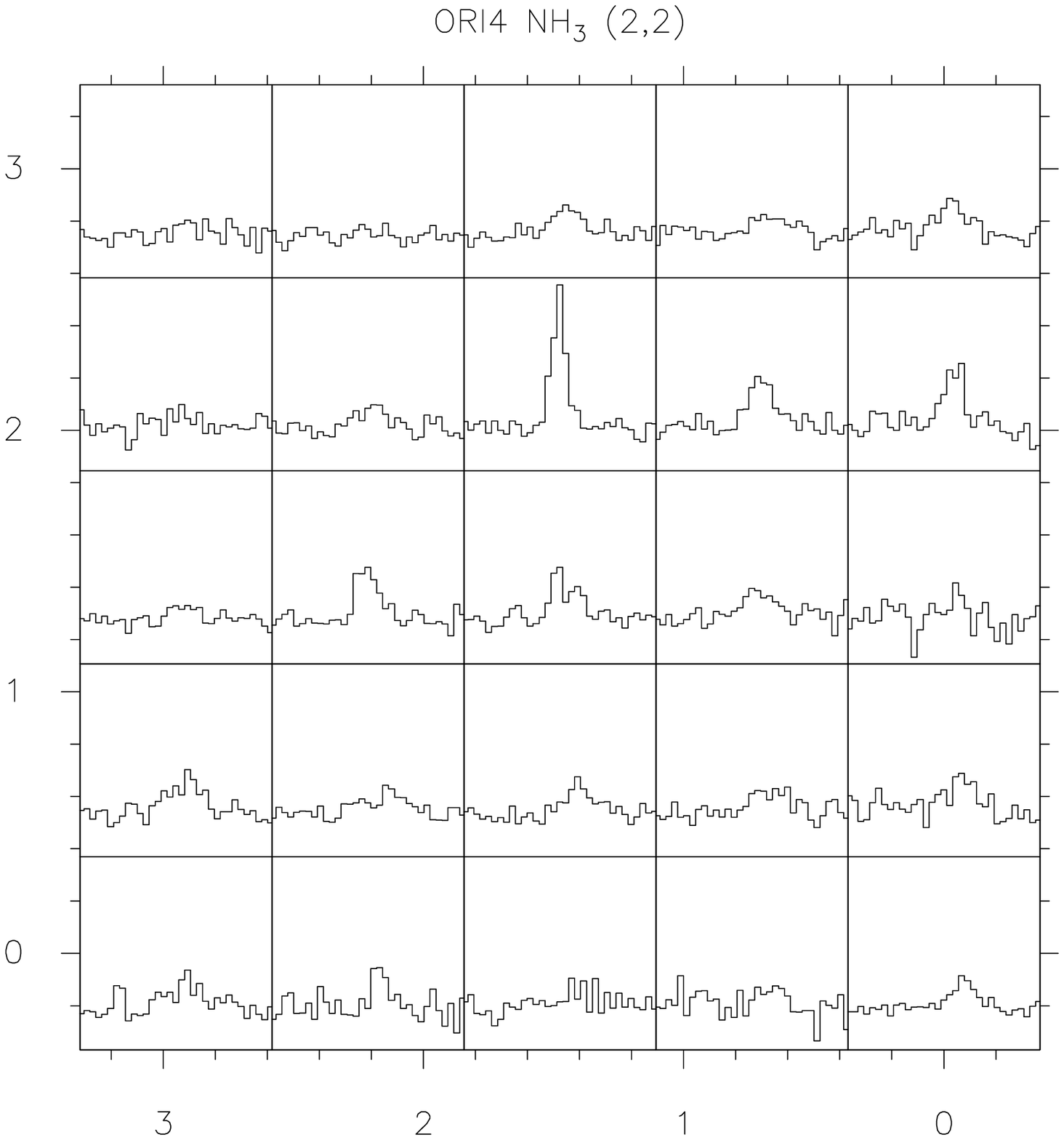}}
   	\caption{Arrays of \ammonia~ spectra of ORI4 at corresponding
spatial offsets, 
   		given in units of minutes of arc
from the central position (Table 1). 
   		Upper panel:  (1,1) spectra plotted within the velocity range -16 to 32 \kms~ 
   		and
the antenna temperature range in the range -0.11 to 0.58 K.  
		Lower panel: (2,2) spectra plotted within the velocity range 4 to 
		12 \kms~ and the antenna temperature in the range -0.06 to 0.21 K.}
      	\label{fig:ori4-specmap}
\end{figure}

The (2,2) line intensity is usually less than 50\% of
the (1,1) intensity. 
Since the two inversion transitions are observed simultaneously,
the signal to noise ratio (S/N) of a (2,2) spectrum is 
a factor of two lower than that of the corresponding (1,1) spectrum.
As shown in the sample spectra map (Figure~\ref{fig:ori4-specmap}),
the limiting factor on obtaining an accurate value of the
kinetic temperature is the noise level in the (2,2) data.

We set a $5\sigma$ criterion, i.e., only those data whose (2,2)
peak $T_A$ is higher than $5\times(RMS~Noise)$ are used in
derivation of kinetic temperature. 
Abiding by this standard, we obtained temperature maps of ORI1, 
ORI2, ORI4 and ORI8 (Figure~\ref{fig:tk}). 
The kinetic temperatures at single points in ORI5 and ORI7 have also been 
determined (see Table 1).
The angular offsets have been converted to distance using a distance of 480
pc to the Orion star molecular cloud \citep{genzel81}.

The kinetic temperatures at positions of 
peak integrated intensities of these sources range from 13 K to 19 K. 
These sources thus do represent a collection of cold, quiescent cloud
cores as hoped for when developing the source selection criteria.

At the centers of these cores, the temperatures are 
generally lower than those toward the edge of the maps. 
This is natural for externally illuminated cores, as the heating from UV is 
reduced by the extinction, which
should result in lower dust temperatures.  
In addition the increased molecular density
expected in these cores provides more cooling for the gas, and lower
gas (kinetic) temperatures should result.
In the Orion molecular cloud, the heating comes mostly from the enhanced
interstellar radiation field, which can explain the
higher temperatures (25 K $\sim$ 30 K) observed outside of these dense
molecular cores.
In the above discussion, there are actually two 
conclusions drawn based on our \ammonia\ observations. 
First, the cores are cold. Second, this temperature variation
is correlated with the distribution of material, in this
case, represented by the integrated intensities.
Are these results credible?
We address the uncertainties of our results in the following Section.

\subsection{Statistical Analysis of Correlations between Temperature and
Intensity \label{stat}}

If the noise distribution is known, we can estimate the
uncertainty of a particular parameter fitted to a spectrum.
In most of our data, the noise in individual channels is
uncorrelated (between channels) Gaussian noise. 
The functions being fitted to the spectra are usually either
polynomials or Gaussians.
Thus, it is easy to obtain the statistical uncertainty of a 
quantity such as the line strength $T_A$.
The uncertainty in a particular parameter is usually given in 
combination with the quality of the data, $\sigma$, where 
$\sigma$ is defined for Gaussian noise
through $P(x) = \frac{1}{\sqrt{2\pi}\sigma} e^{-x^2/(2\sigma^2)}$
with $P(x)$ being the probability density function of the noise amplitude $x$
and the mean noise amplitude being zero.
$\sigma$ is also called the RMS (root mean square) variation.
The full width to half maximum (FWHM) of the Gaussian
function is $\sqrt{8\ln{2} } \sigma$.  
Subtracting the fitted function from the data, we measure the $\sigma$ from the
residual noise so that we can make a statement about the line strength, 
such as $T_A$ being at 3$\sigma$ level.

Such statements, however, do not usually have 
direct physical meaning.  
For example, as discussed in the previous sections on deriving the kinetic temperature, 
two fitted quantities, one from a \ammonia\ (1,1) spectrum and the 
other from a \ammonia\ (2,2) spectrum,
are combined to derive the rotational temperature. 
Then, through excitation calculations, 
the rotational temperature is converted to the kinetic temperature.
What concerns us is the level of uncertainty in the kinetic
temperature resulting from the noise in the original 
(1,1) and (2,2) spectra. 
An analytic solution through error propagation is not practical since 
the derivation involves integration, interpolation, and complicated functions.

We have devised an empirical approach to determine the uncertainty in the
kinetic temperature, which is based on adding random
noise to a perfect spectrum and using it to derive 
the kinetic temperature. 
By repeating this experiment many times, each with independently 
generated noise, we obtain an ensemble of spectra, each of which 
is associated with a specific $T_k$ through the fitting procedure.
Since the added noise in each spectrum is independent, each 
determined $T_k$ can be treated as an independent representation of a
 random variable. 
According to the central limit theorem, 
a Gaussian describes the distribution of kinetic temperatures 
regardless of the intrinsic distribution of $T_k$ with respect 
to noise magnitude.

Figure~\ref{fig:noiseadd} illustrates how noisy
spectra are constructed by adding Gaussian noise 
(Figure~\ref{fig:noise}) to a `perfect' spectrum. The
`perfect' spectra are obtained by averaging over our maps,
and thus have negligible noise. The RMS level of 0.01 K
is a typical value achieved in our observations at
Effelsberg. `Perfect' spectra are so scaled that
the (2,2) peak is at a certain signal to noise ratio,
e.g. $5\sigma$ or $10\sigma$.

The uncertainty in $T_k$ depends not only on the 
noise $\sigma$, but also on the signal strength. A better fit
with less spread can be achieved at a higher signal to noise 
ratio. The results of our `noise' experiment 
(Figure~\ref{fig:sigma}) are consistent with this expectation.
The kinetic temperatures obtained through the simulation 
are consistent with a Gaussian distribution.
At the $5\sigma$ level, which is our criterion for attempting
any temperature calculation, the RMS of the $T_k$ distribution
is 1.8 K. 
At the $10\sigma$ level, which is a typical S/N for 
data around the cloud centers, the RMS is 0.9 K. 
In between, the RMS of the derived $T_k$ goes down roughly linearly
as a function of the S/N (Figure~\ref{fig:5to10sigma}). 
The dependence of the RMS of the fitted $T_k$ on the S/N
ratio of the original spectra is well described 
by
\begin{equation}
	RMS (\text{K}) = -0.15+\frac{10.3}{S/N}+0.01(S/N) \lc 
\label{sn}
\end{equation}
which is used to assign uncertainties to individual
data points when needed. This numerical fit is applicable only
to a reasonable range (3 to 20)
of the signal to noise ratio. 

Such a noise distribution, obtained through appropriate simulations, 
enables us to make meaningful statements about kinetic temperatures.
For example, at the center of ORI1, $T_k = 19\pm$0.9 K at a 67\% 
confidence level and at the edge of ORI1, $T_k=25\pm 1.8$ K at
a 67\% confidence level. 
With the uncertainties of the derived $T_k$ known, we can
tackle the problem of assessing the validity of the temperature
gradients and their dependence on core column density, as discussed
earlier. 

\begin{figure}[htp]
      	\centering 
     	\subfigure{\includegraphics[width=8cm]{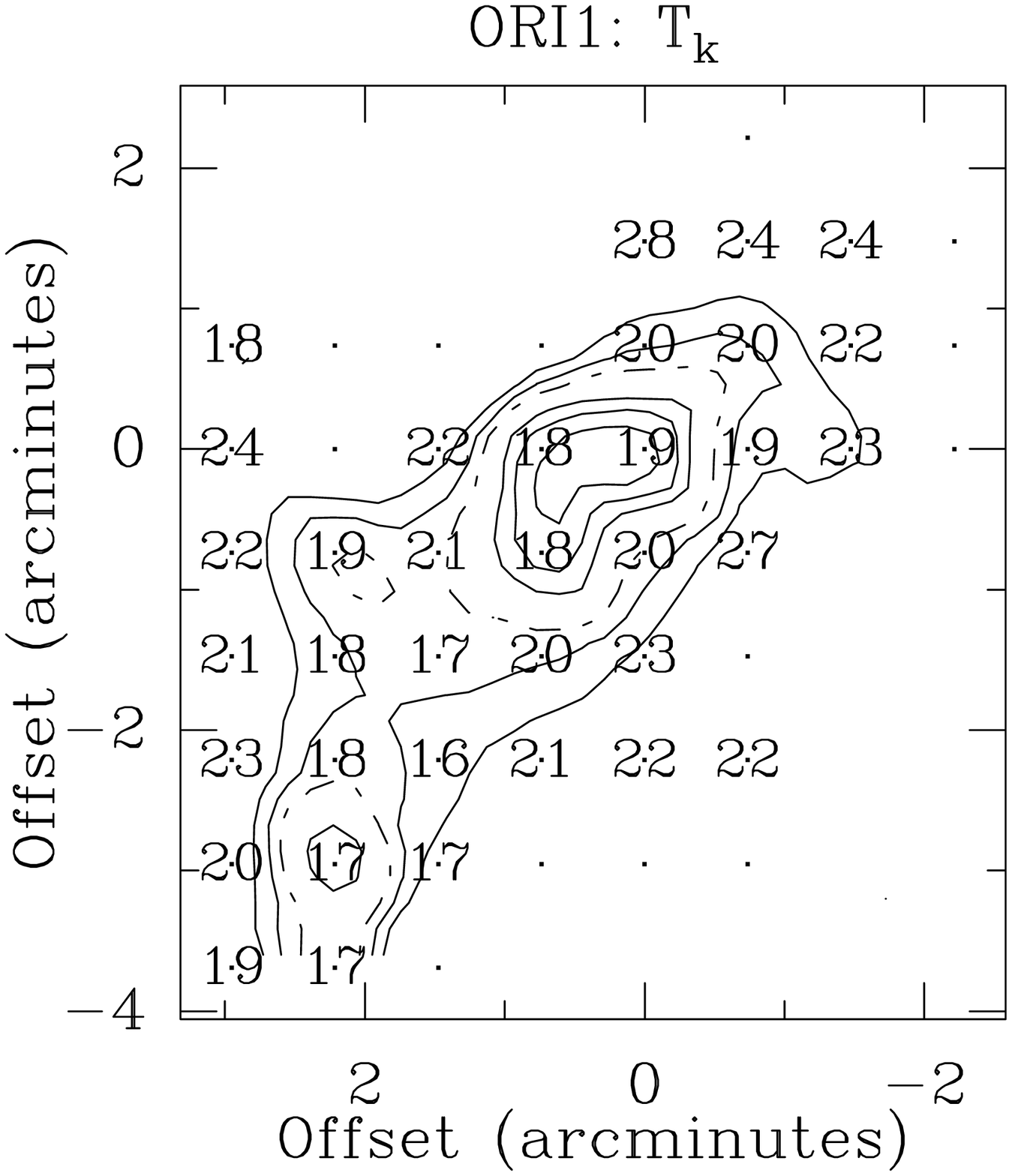}}
   	\subfigure{\includegraphics[width=8cm]{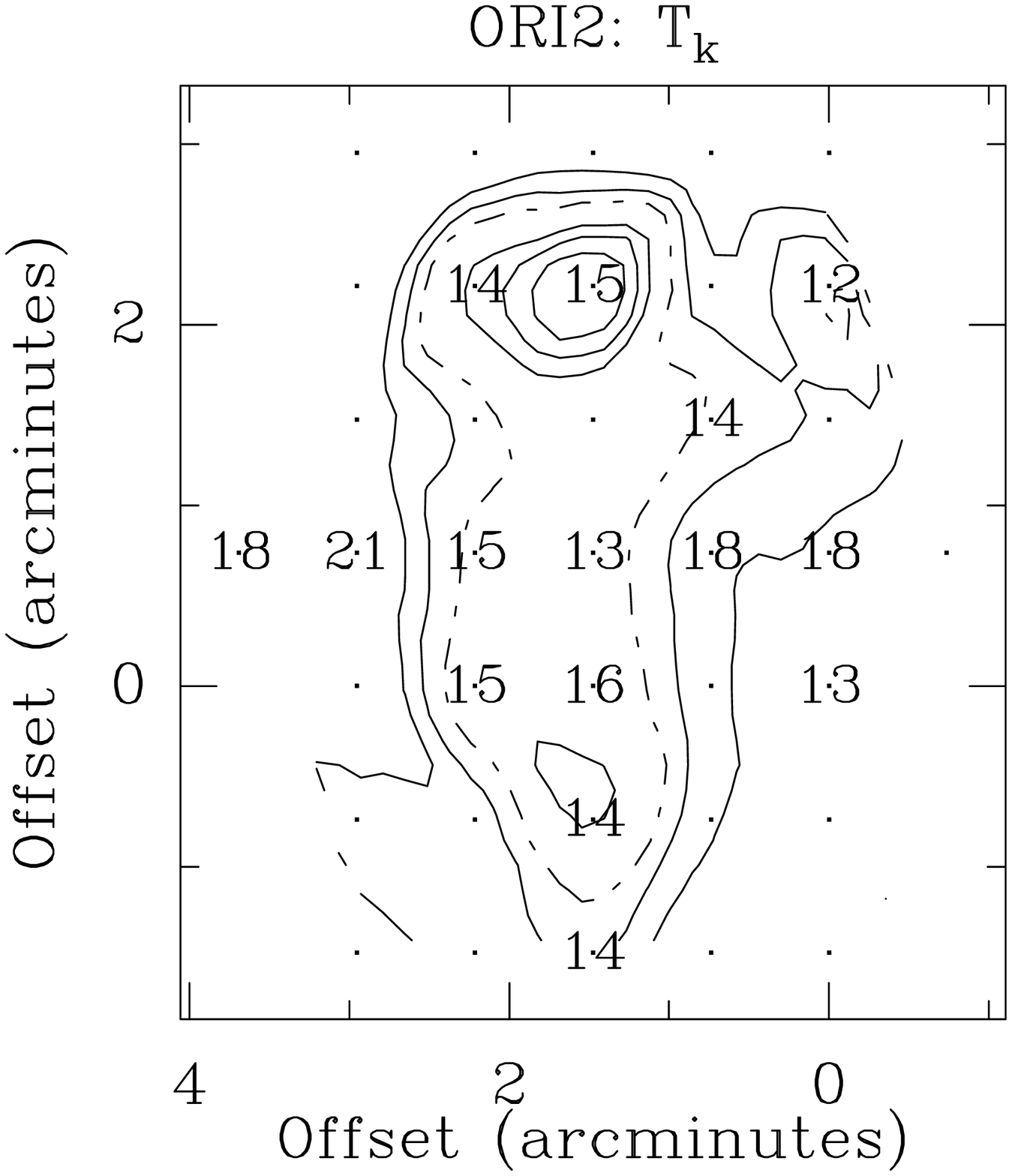}}\\
     	\subfigure{\includegraphics[width=8cm]{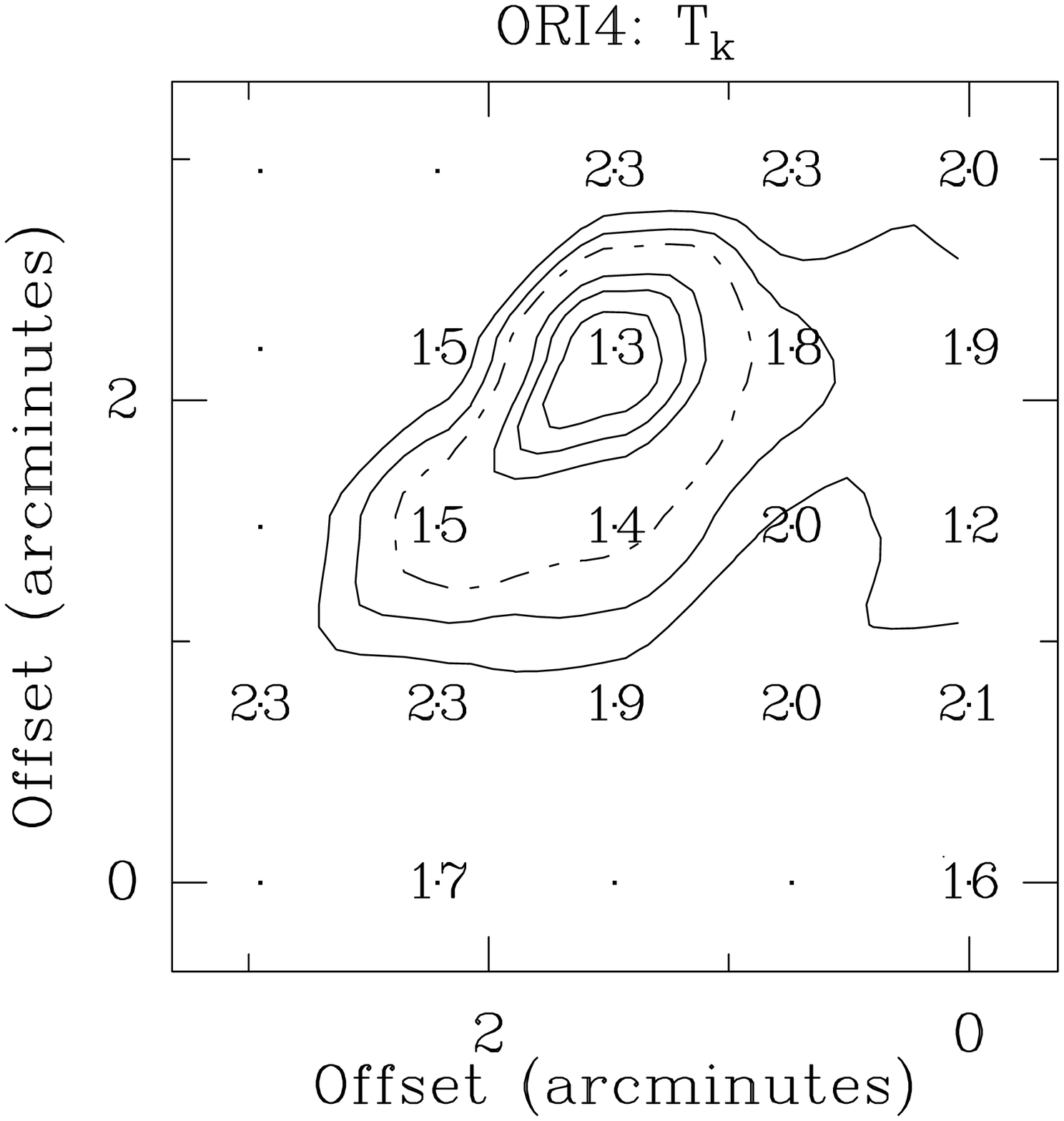}}    	
       	\subfigure{\includegraphics[width=8cm]{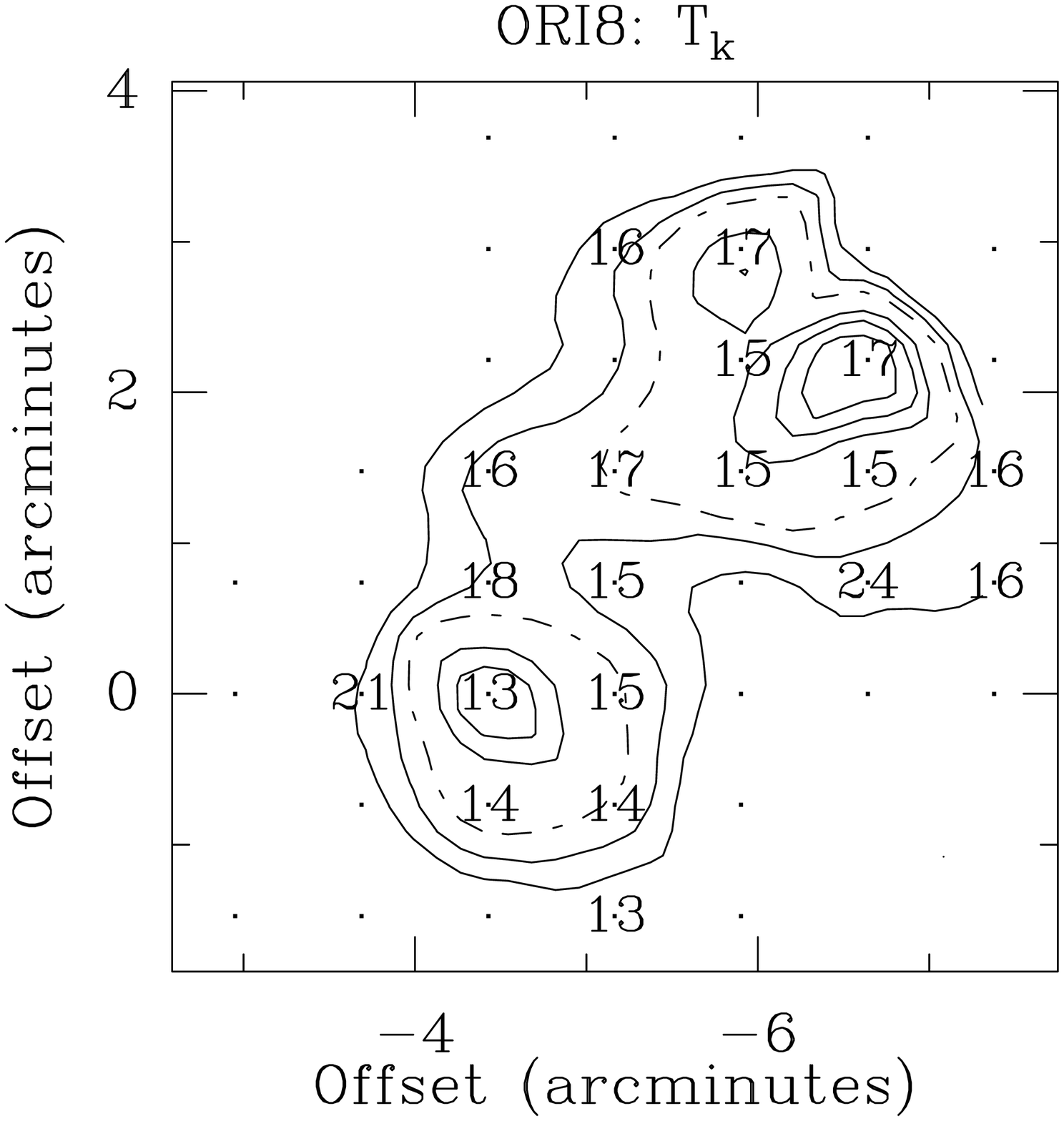}}
	\caption{Derived kinetic temperatures overlaid with contours of the
		integrated intensity of the \ammonia~(1,1) transition. The offset is given 
		in arcminutes
relative to center positions given in Table 1.
		The contour levels are at 0.3, 0.4, 0.5, 0.7, 0.8, and 0.9 times the peak value. 
		The 0.5 contour is indicated by dashed lines. 
		The dots indicate the positions where data have been obtained.}
	\label{fig:tk}   
\end{figure}

\begin{figure}[htp]
      \centering 
     	\subfigure{\includegraphics[width=6.5cm]{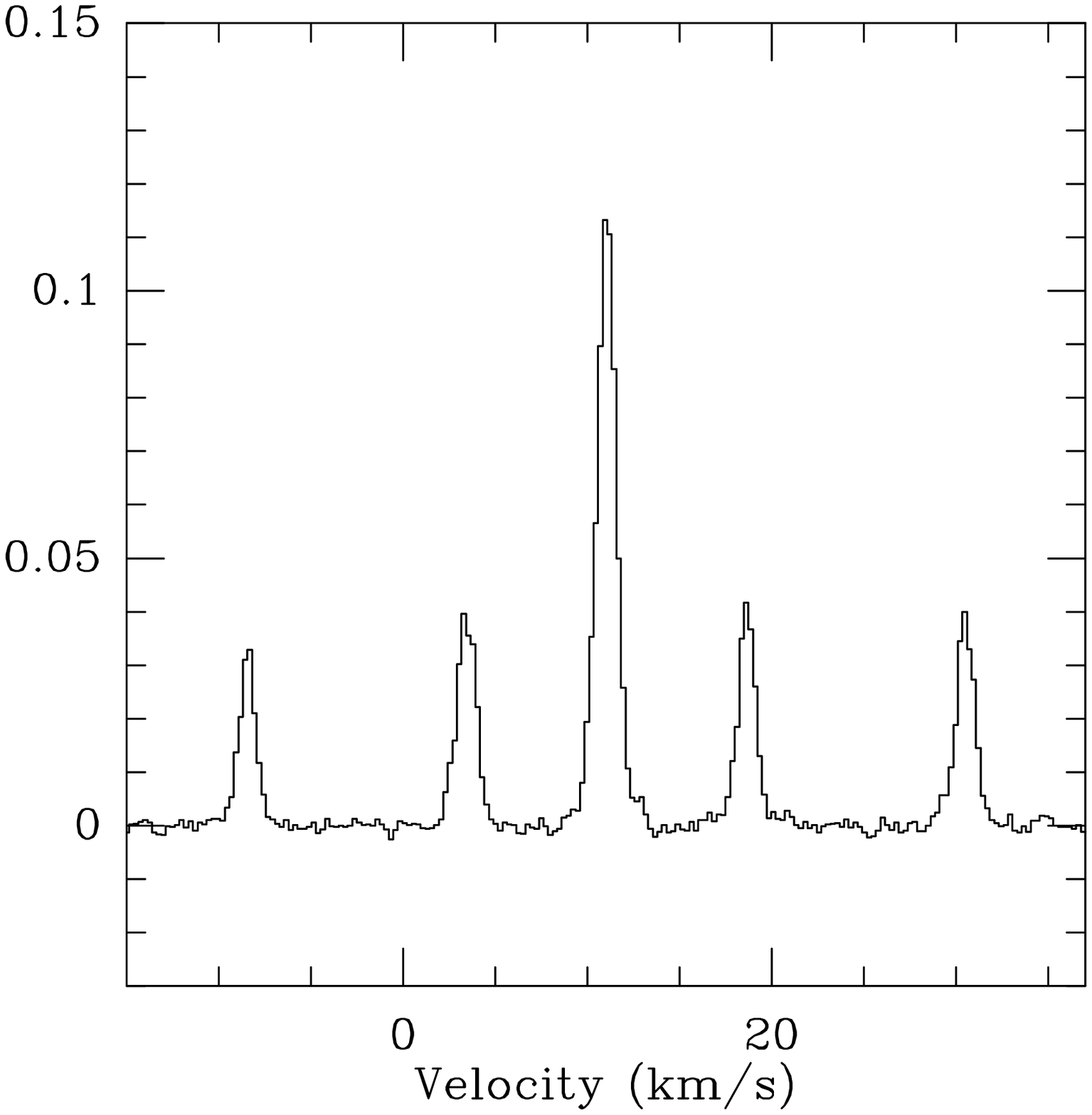}}
     	\subfigure{\includegraphics[width=6.5cm]{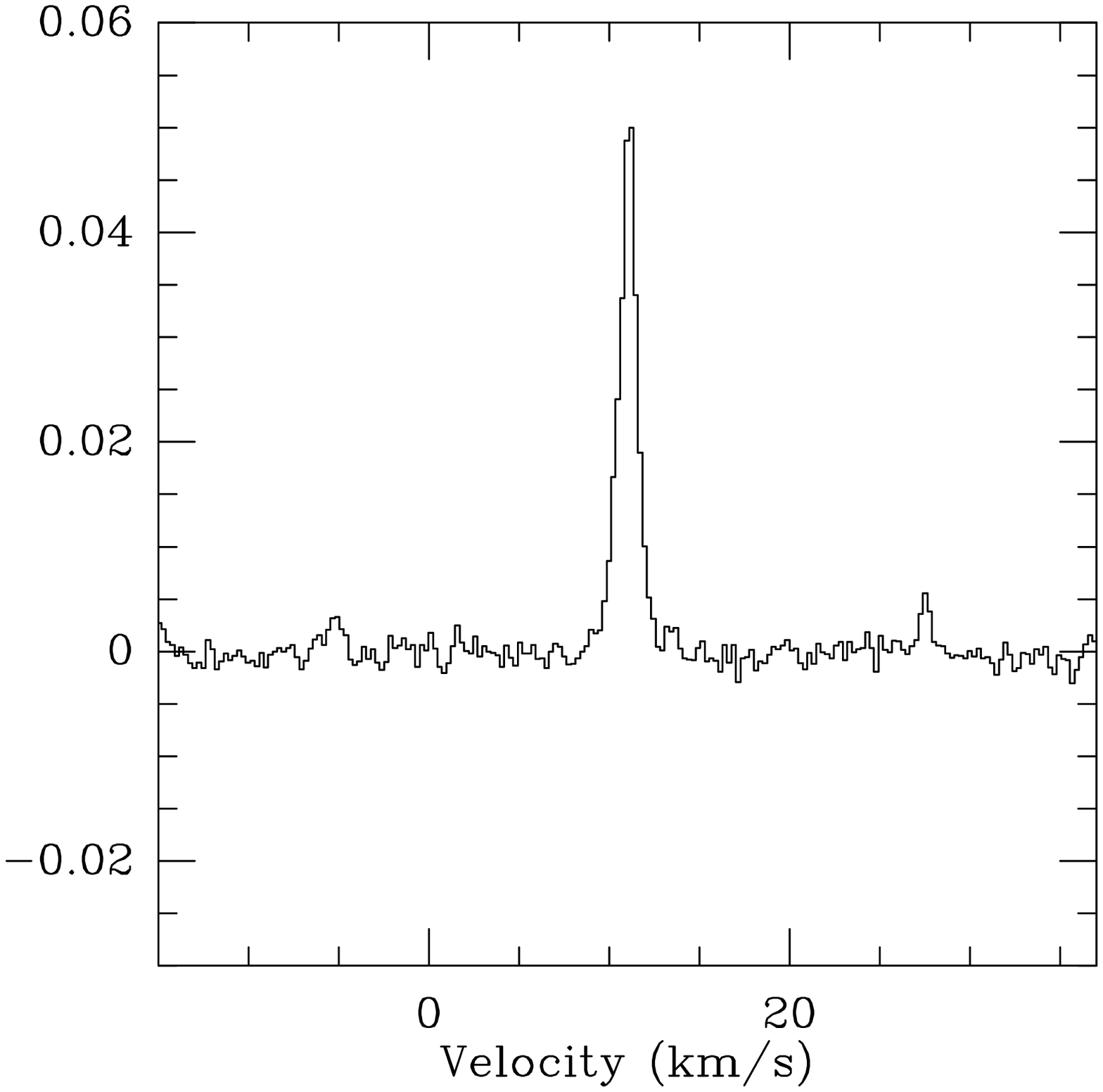}}\\
     	\subfigure{\includegraphics[width=6.5cm]{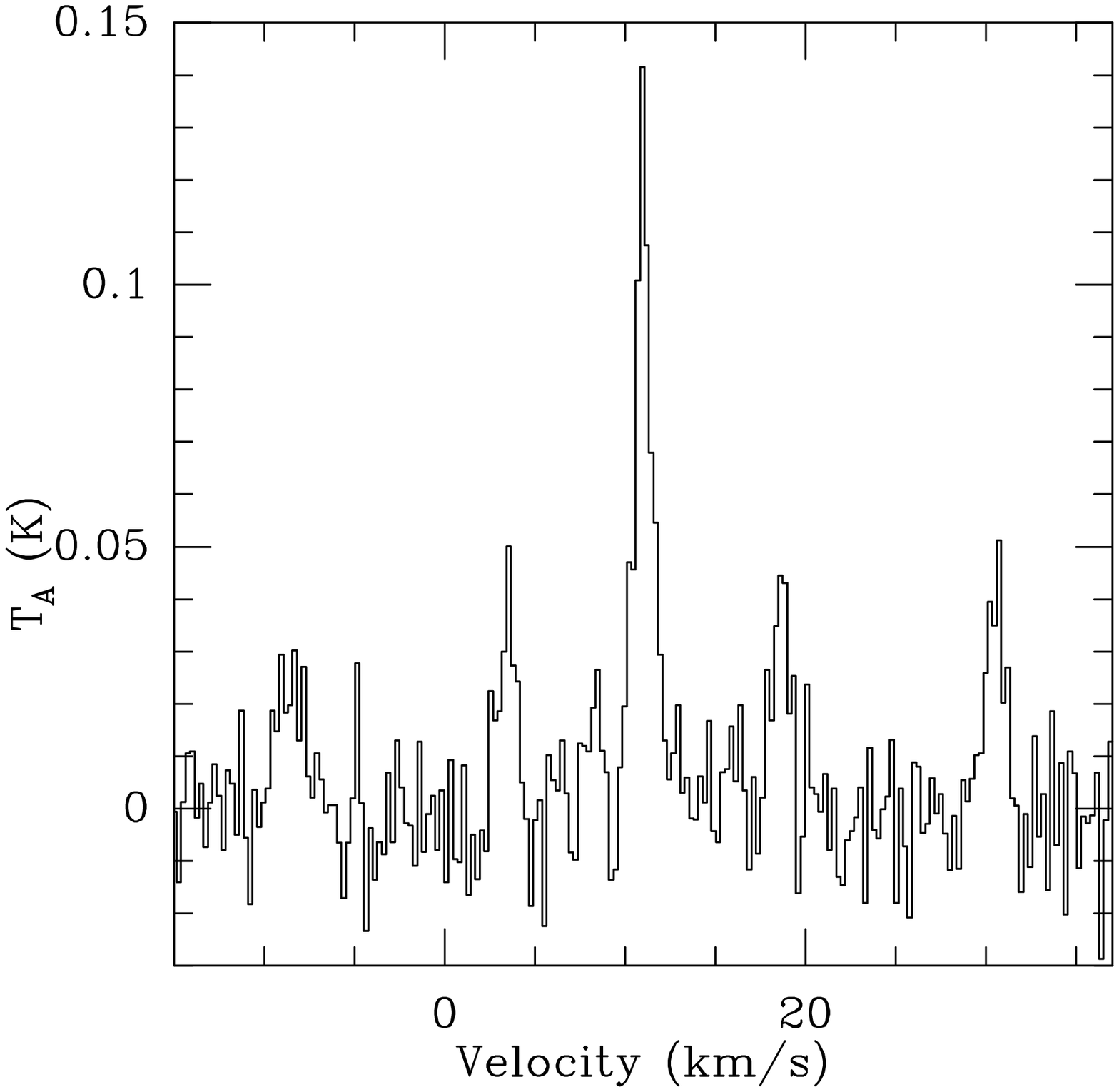}}
     	\subfigure{\includegraphics[width=6.5cm]{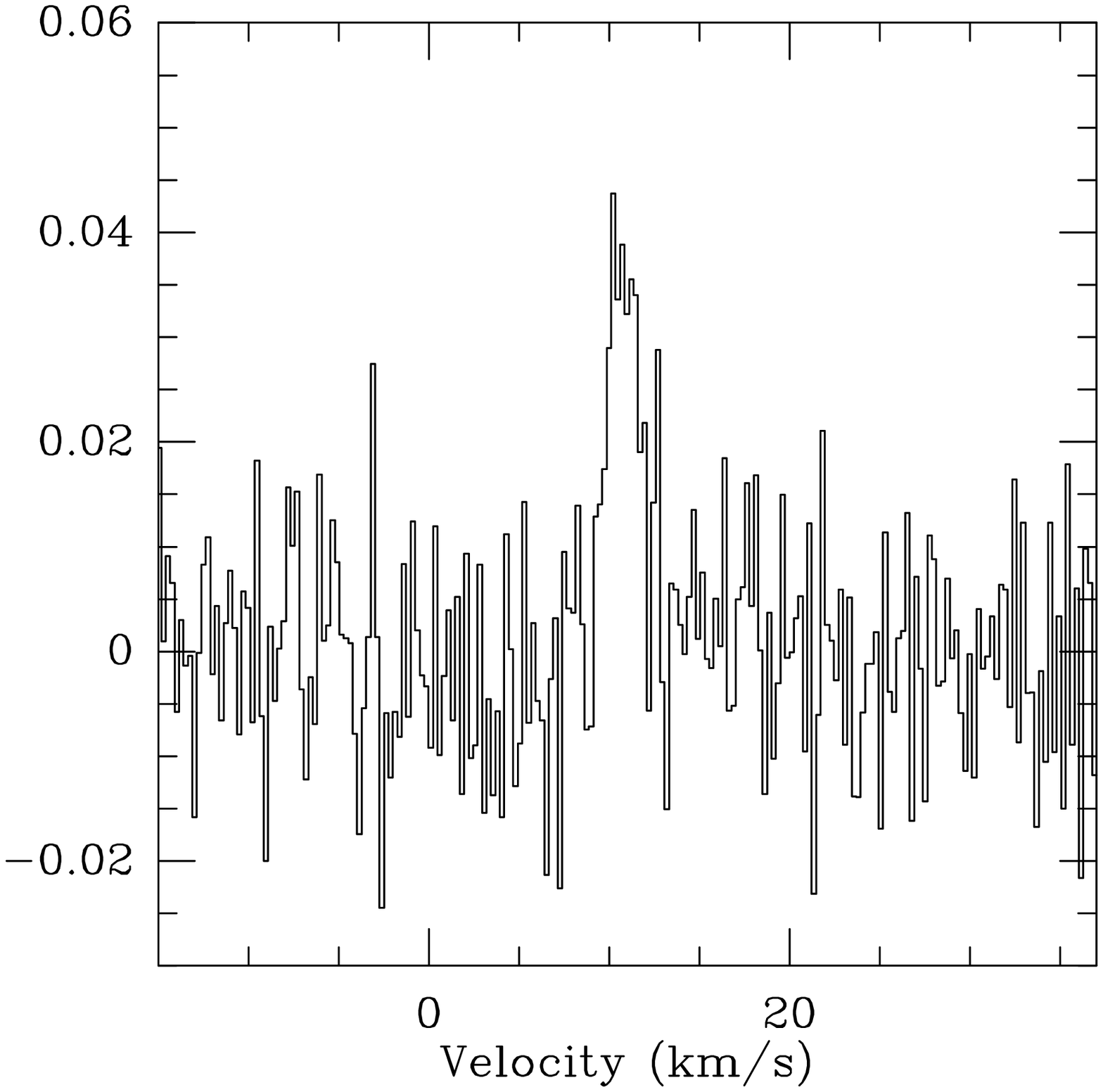}}
     	\caption{`Perfect' spectra of \ammonia\ (upper panels) with
		added noise (lower panels). The (1,1) spectra are shown on the
		left and the (2,2) spectra on the right.
		The noise is Gaussian with 0.01 K RMS, which 
		makes the Gaussian fit to the (2,2) spectrum significant 
		at about the 5 $\sigma$ level.}
     	\label{fig:noiseadd}
\end{figure}

\begin{figure}[htp]
      	\centering 
      	\includegraphics[width=.60\textwidth, angle=90]{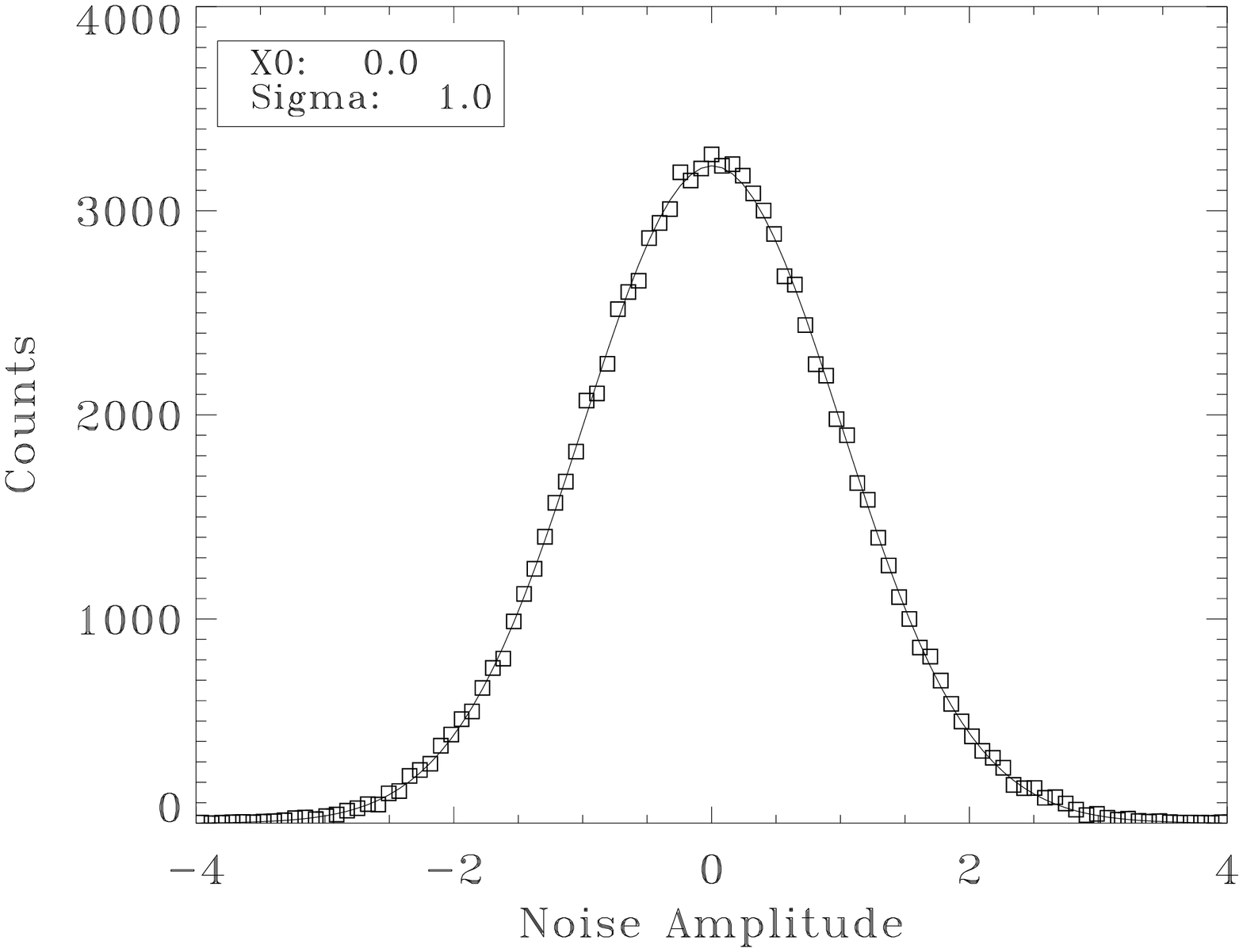} \\
      	\includegraphics[width=.58\textwidth, angle =270]{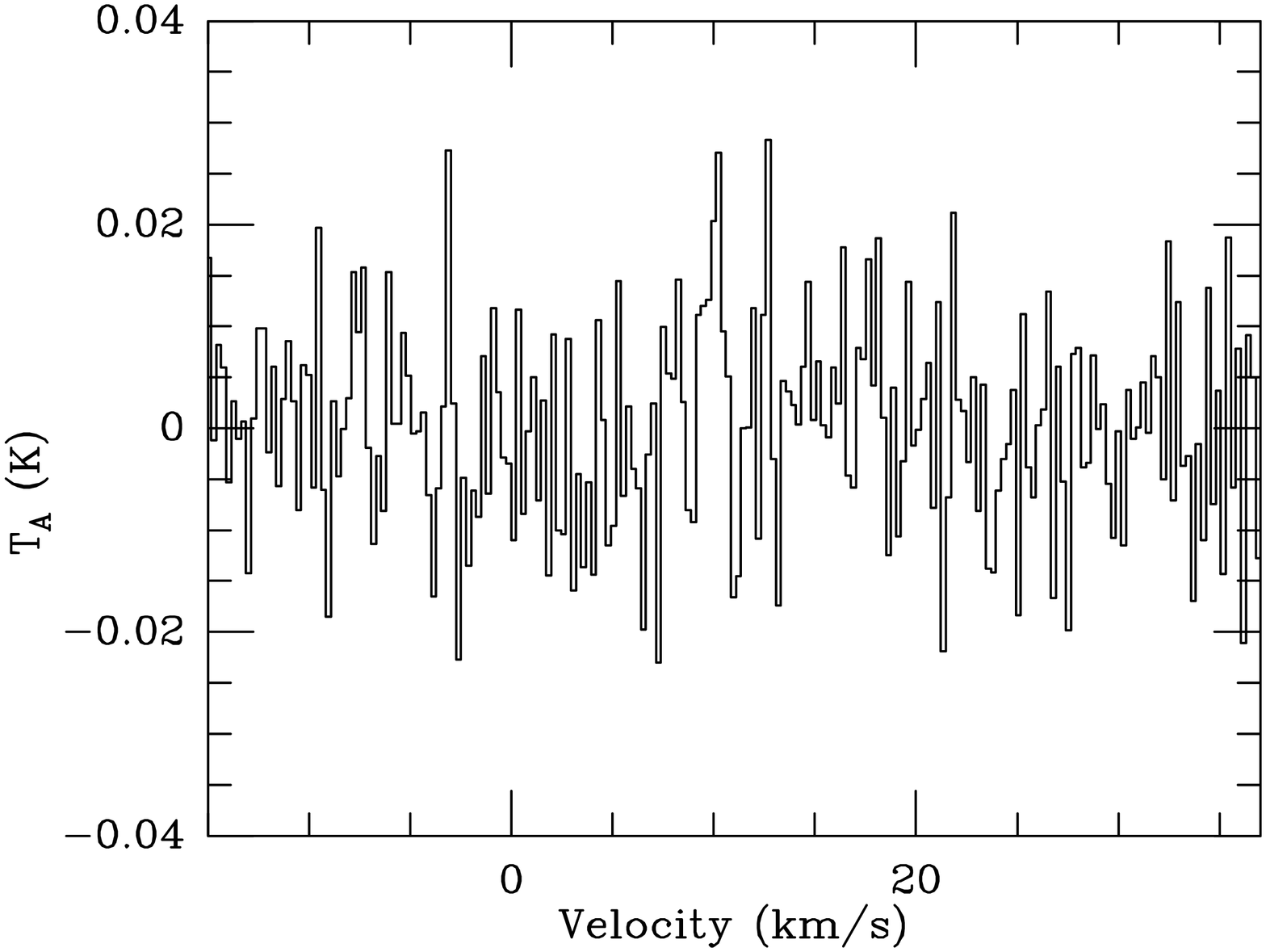} 
     	\caption{Distribution of a Gaussian random variable with zero mean 
		(X0=0) and unit variance (Sigma=1). 
		In the upper panel, the squares indicates 
		counts of generated values of the variable in each bin and 
		the line indicates the fitted Gaussian. The lower panel is 
		an example of such noise, with the RMS scaled to 0.01 K.}
     	\label{fig:noise}
\end{figure}

\begin{figure}[htp]
      	\centering 
      	\includegraphics[width=.60\textwidth, angle=90]{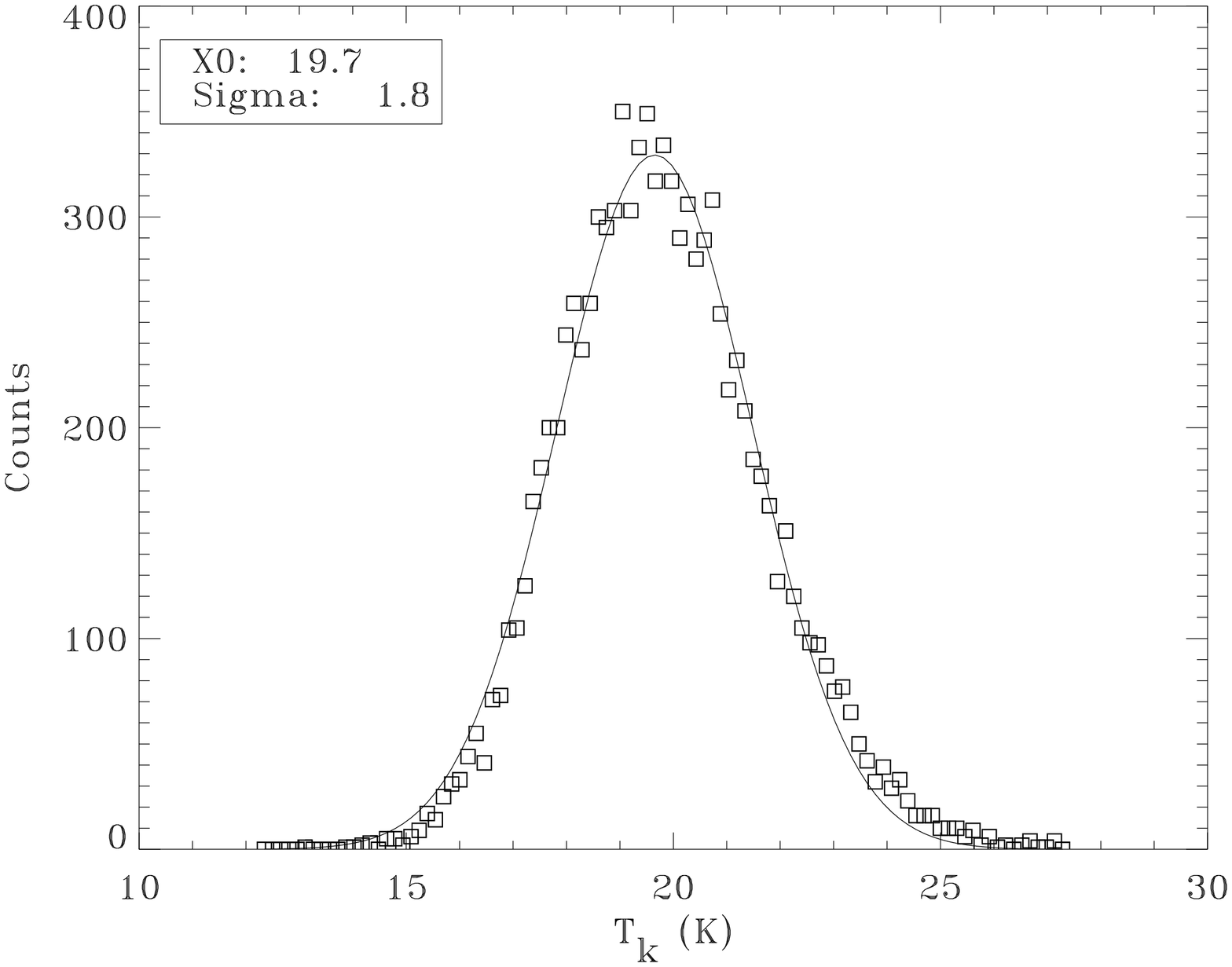} \\
      	\includegraphics[width=.60\textwidth, angle =90]{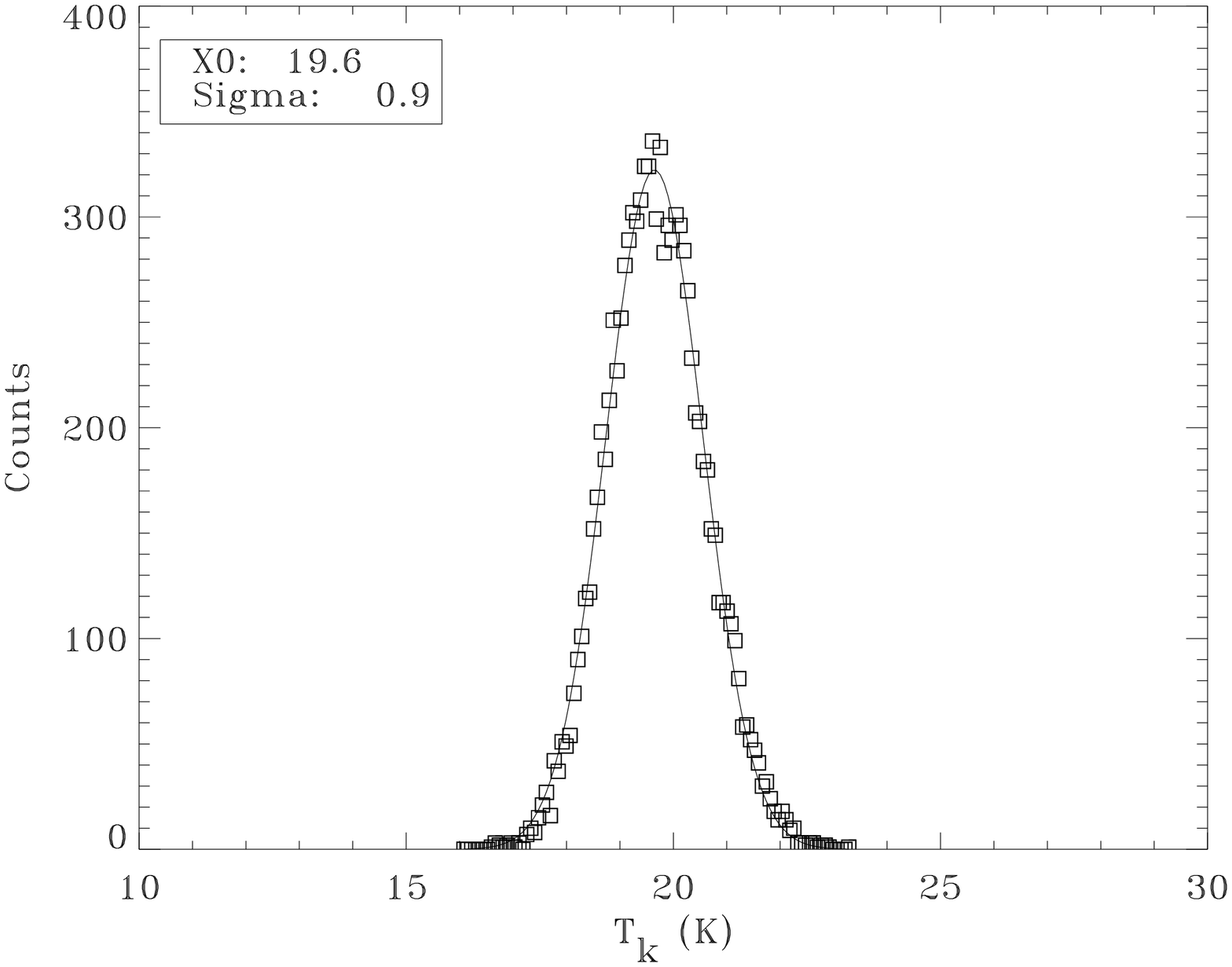} 
     	\caption{Distribution of kinetic temperatures
		obtained from artificially generated noisy \ammonia\ spectra 
		whose RMS noise is 0.01 K. 10,000 pairs of noisy 
		spectra were generated and used to calculate $T_k$. 
		X0 and Sigma give the mean and the variance, respectively,
		of the samples of the derived kinetic temperatures.
		The kinetic temperature was 20 K in all cases. 
		Upper panel:  $5\sigma$ signal. Lower panel: $10\sigma$ signal.}
	\label{fig:sigma}
\end{figure}

\begin{figure}[htp]
      	\centering 
      	\includegraphics[width=.80\textwidth]{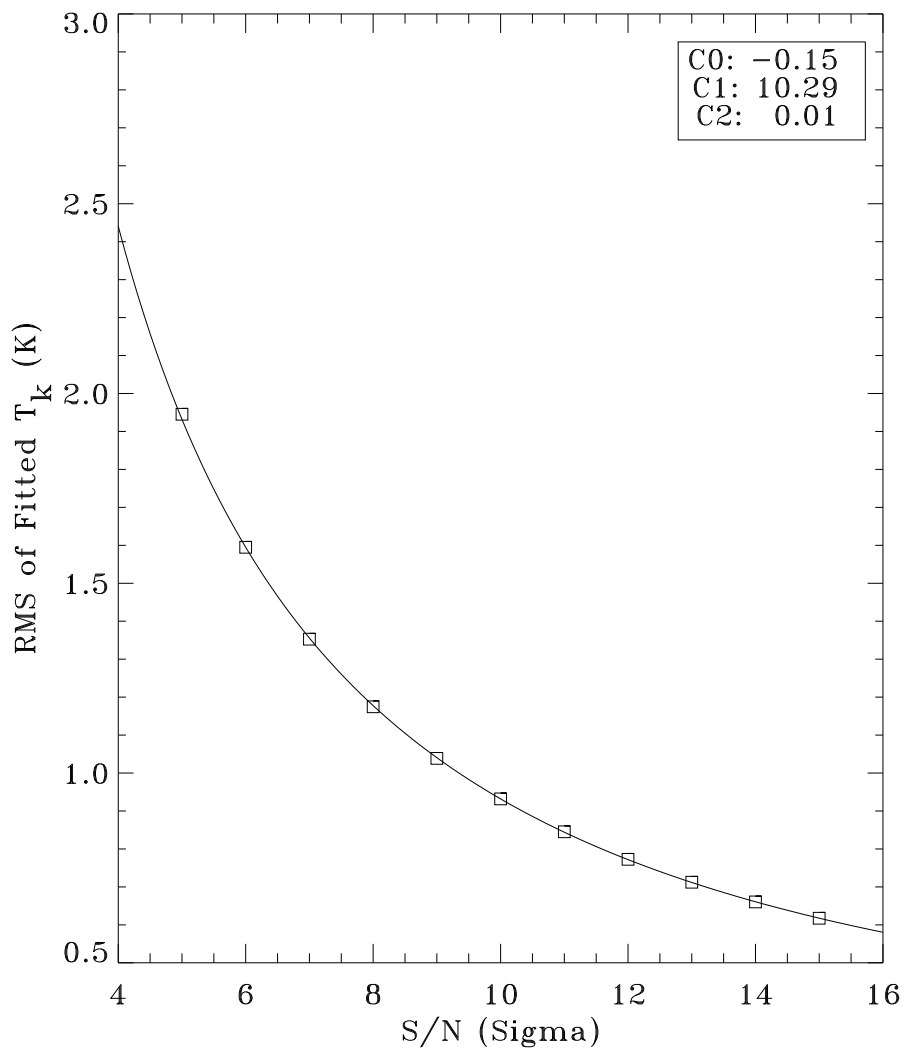}
     	\caption{Uncertainty of the derived $T_k$ vs.\ the 
		S/N ratio of generated \ammonia\ spectra. Each point is based
		on 1000 pairs of (1,1) and (2,2) spectra with 0.01 K RMS noise 
		and specified S/N ratio for the (2,2) line.  
		The relative weakness of this line (found observationally) results in its S/N 
		ratio being the determinant of $T_k$; the S/N ratio of the (1,1) line is 
		accordingly set to infinity.
		The solid line indicates the fitted
function, $C_0+C_1/(S/N)+C_2(S/N)$, 
		with the three 
coefficients given in the box.}
	\label{fig:5to10sigma}
\end{figure}

First, we examine the claim that there are significant
temperature variations within the maps. 
The mapping data of $T_k$ (Figure~\ref{fig:tk}) 
can be divided into two groups according to the integrated intensity 
at each point.
The central group includes those data points whose intensity
is higher than half of the peak value of each map, i.e., the points
within the dotted contour line of integrated intensity in Figure~\ref{fig:tk}. 
The edge group includes all other points. 

The mean $T_k$ in each map is subtracted from the data points in our sample. 
Thus, we have four maps of temperature variations, all with zero mean. 
We can combine them into one combined sample including 85 points, out of which
54 are in the edge group and 31 are in the central group.
The mean of edge group is 0.76 K and that of the central group
is -1.3 K. 
On average, the temperature in central regions
is 2.1 Kelvin lower than that on the outside. 

To evaluate the significance regarding the difference of
means ($\overline{A}$ and $\overline{B}$) of two samples ($A$ and $B$), 
we use Student's t test, whose relevant statistic is
\begin{equation}
	t=\frac{\overline{A}-\overline{B}}{( \Delta A/N_A+\Delta B/N_B) ^{1/2}} \lc
\label{st}
\end{equation}
where $\Delta A$ is the variance of sample $A$, $N_A$ is the number
of points in sample $A$, and $\Delta B$ and $N_B$ are these
quantities for sample $B$. 
If the distribution is known to be a Gaussian, $\Delta$
is equal to $\sigma^2$.
In our case, the variance is 0.81 K and 3.2 K for the central group 
and the edge group, respectively. 
Note that the two groups have different
variances due to different S/N of the \ammonia\ spectra.
Using Student's t test on samples with different variances
may be troubling, but not for cases where both distributions
are known to be Gaussian.

The number of degrees of freedom for Student's t
distribution in this problem is
\begin{equation}
	\nu=\frac{(\Delta A/N_A+\Delta B/N_B) ^2}
		{ \frac{(\Delta A/N_A)^2}{N_A-1}+ \frac{(\Delta B/N_B)^2}{N_B-1} } \lp
\label{nu}
\end{equation}

For two samples having equal mean values,
the probability with which the difference in means would
be just this large or larger by chance, is calculated
from Student's distribution using
\begin{equation}
p(null)=I_{\frac{\nu}{\nu+t^2}}(\nu/2, 1/2) \lc
\label{atu}
\end{equation}
where $I$ is the incomplete beta function \citep*[see e.g.\ ][]{nr}.
Applying equations~\eqref{st}, \eqref{nu} and \eqref{atu},
we obtain $p(null)$ to be $1.1\times10^{-9}$. 
The null hypothesis is very unlikely:  the temperatures in central regions
of the dense cores studied here are statistically significantly 
lower than in the outer regions.

\begin{figure}[htp]
      	\centering 
     	\subfigure{\includegraphics[width=8cm, angle=90]{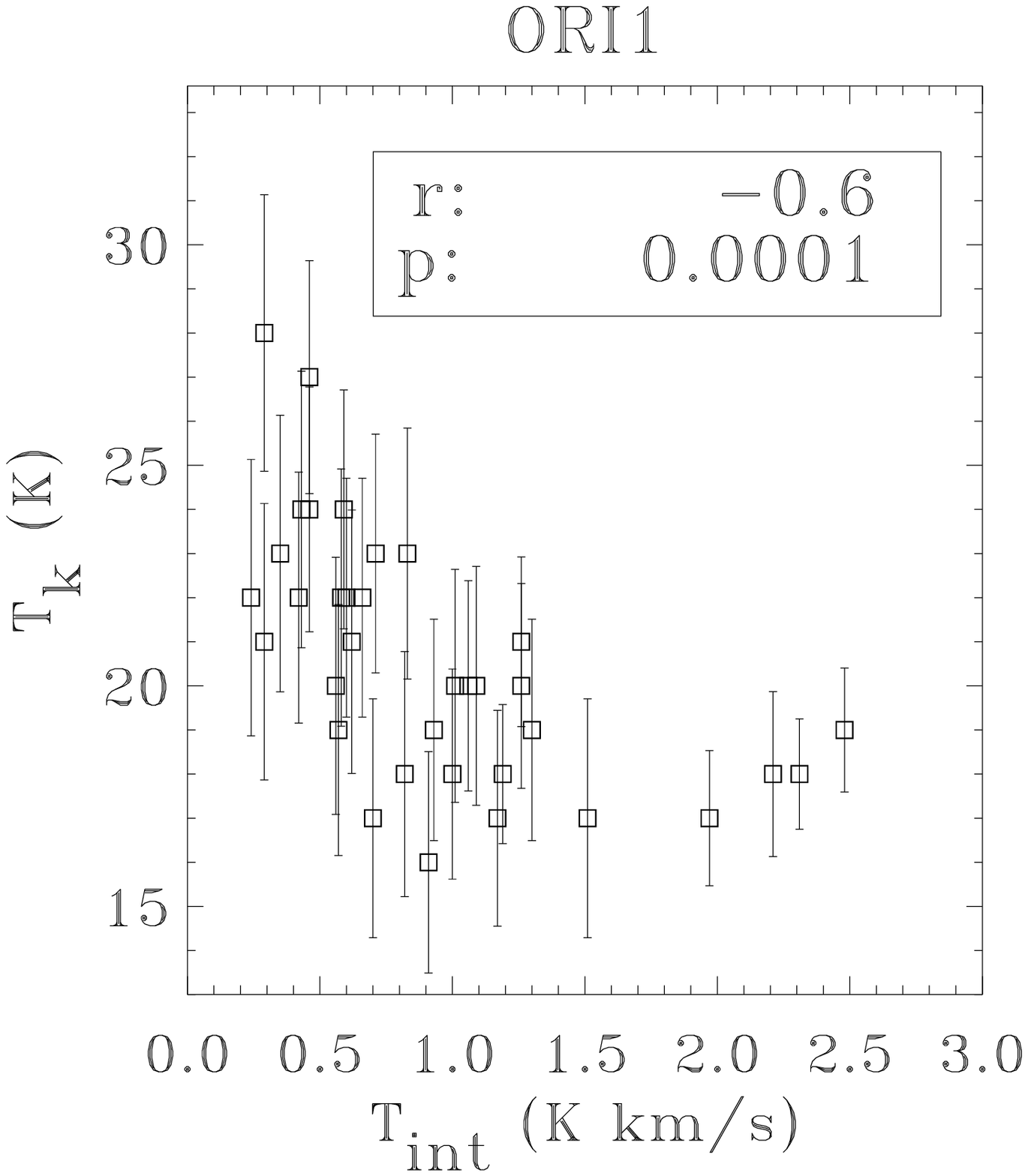}}
     	\subfigure{\includegraphics[width=8cm, angle=90]{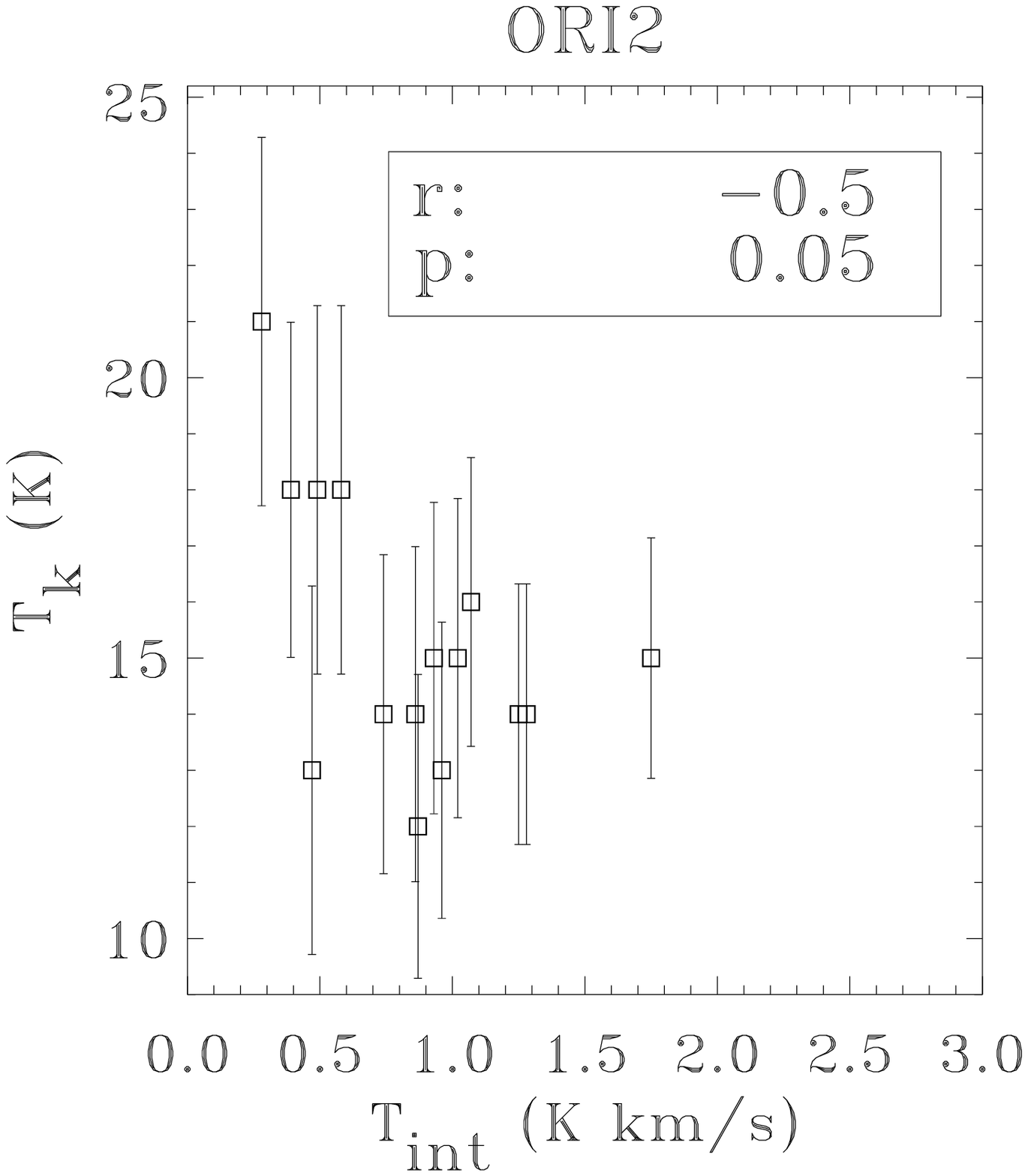}}\\
 	\subfigure{\includegraphics[width=8cm, angle=90]{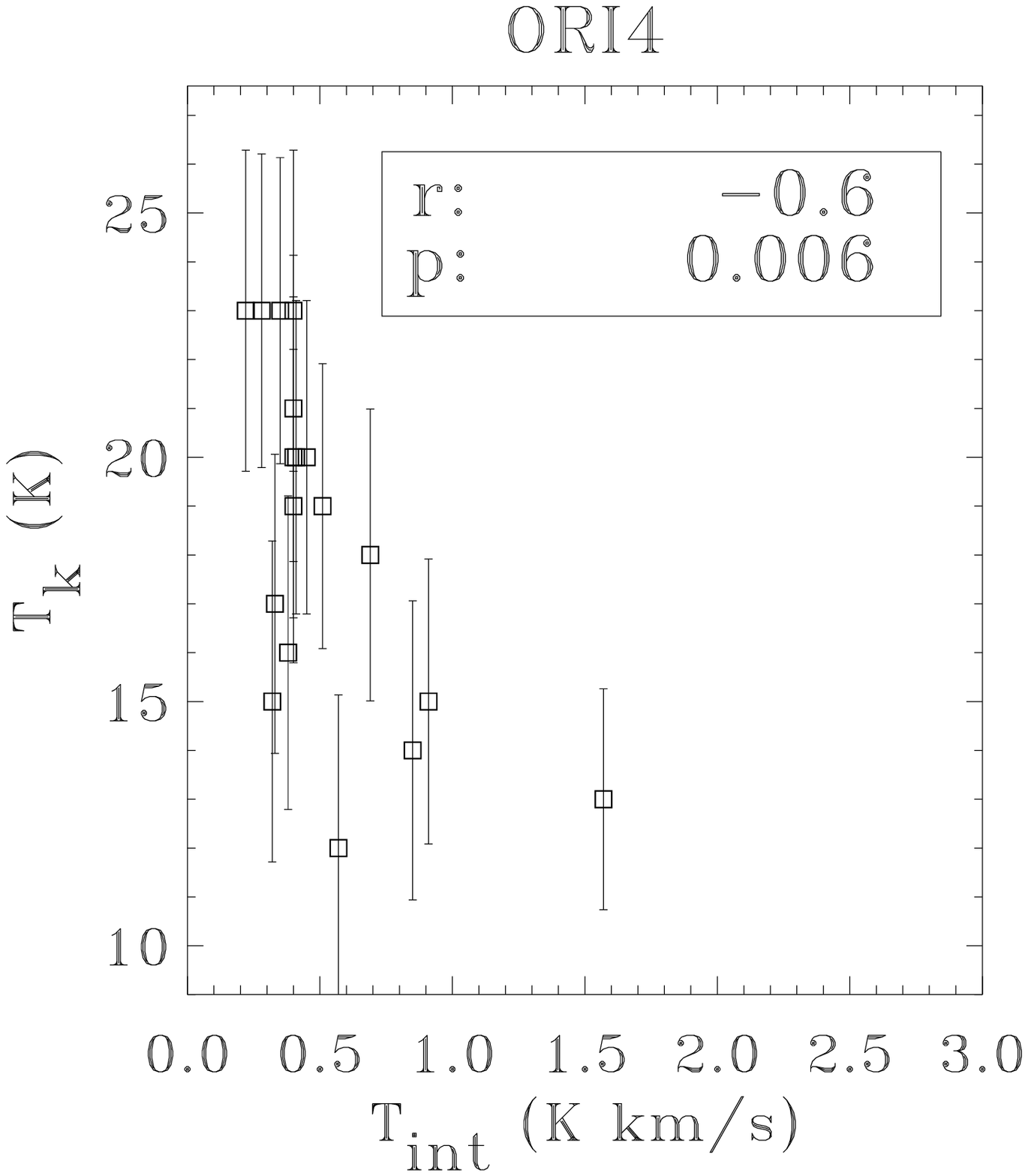}}
     	\subfigure{\includegraphics[width=8cm, angle=90]{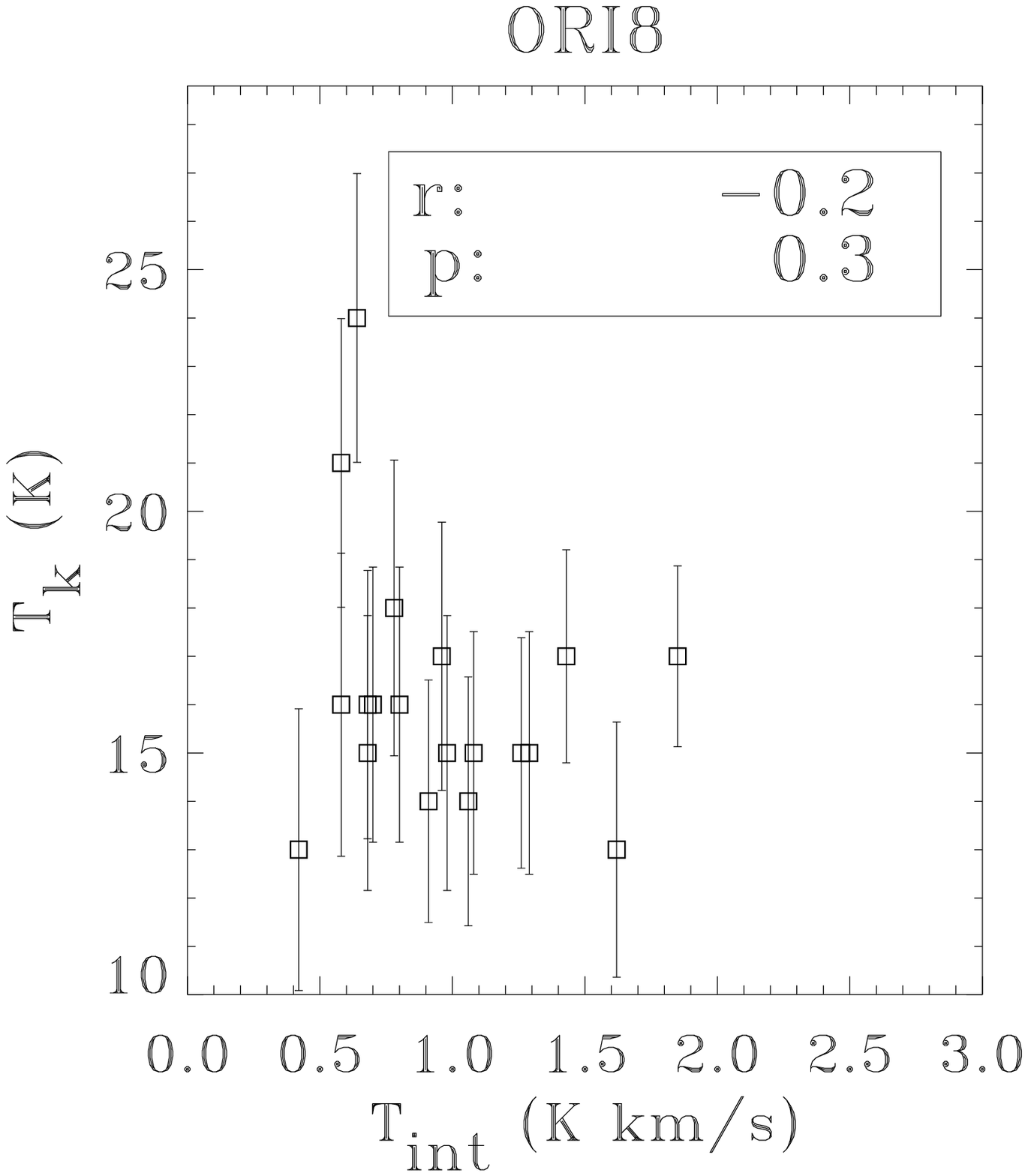}} 
    	\caption{Correlations between integrated intensities of \ammonia\ (1,1)
		line and the derived $T_k$. The Pearson correlation coefficient 
		$r$ and the significance of the null hypothesis $p$ 
		(no correlation) are given in each small box. The error bars are 
		at $\pm$1$\sigma$ level, given by Eq.~\eqref{sn}.} 
    	\label{fig:corr}
\end{figure}

Now let us consider the second claim raised in the previous section concerning
the correlation between intensity and temperature. 
For pairs of quantities $(T_k^i, T^i_{int})$, $i=1,2,\dots, N$, 
the most commonly used linear correlation coefficient is called Pearson's r, 
given by 
\begin{equation}
r=\frac{\sum_i (T^i_k-\overline{T}_k)(T^i_{int}-\overline{T}_{int})}
		{\sqrt{\sum_i(T^i_k-\overline{T_k})^2}
		\sqrt{\sum_i(T^i_{int}-\overline{T}_{int})^2} } \lp
\end{equation}
$r$ ranges from -1 to 1 and indicates the level of correlation.
But more important is the confidence level of any derived $r$.
Even when the number of data pairs is not very large, as is the case here,
the statistic
\begin{equation}
	t=r\sqrt{ \frac{N-2}{1-r^2}} 
\end{equation}
is still distributed like Student's t-distribution in the null case
(no correlation). 
The significance level of the null hypothesis is
thus given by $p(null)$ (Eq.~\ref{atu}).
We calculate $r$ and $p$ for the four temperature/intensity maps 
of Orion cores. 
The results are given in Figure~\ref{fig:corr}.

For ORI1, the anti--correlation between $T_k$ and
$T_{int}$ is clear in the sense that the probability for two 
uncorrelated samples producing the same $r$ by chance is
as small as 0.01\%. 
A similar statement can be made for ORI2 and ORI4. 
In ORI8, the probability of 30\% is not very
small for the null hypothesis. 
This is probably due to the combination of the larger size of the core 
with the limited size of our \ammonia\ map, which does not extend 
far enough to include outer regions with higher $T_k$.

\subsection{Turbulence \label{tub}}
\begin{figure}[htp]
      	\centering 
     	\subfigure{\includegraphics[width=8cm]{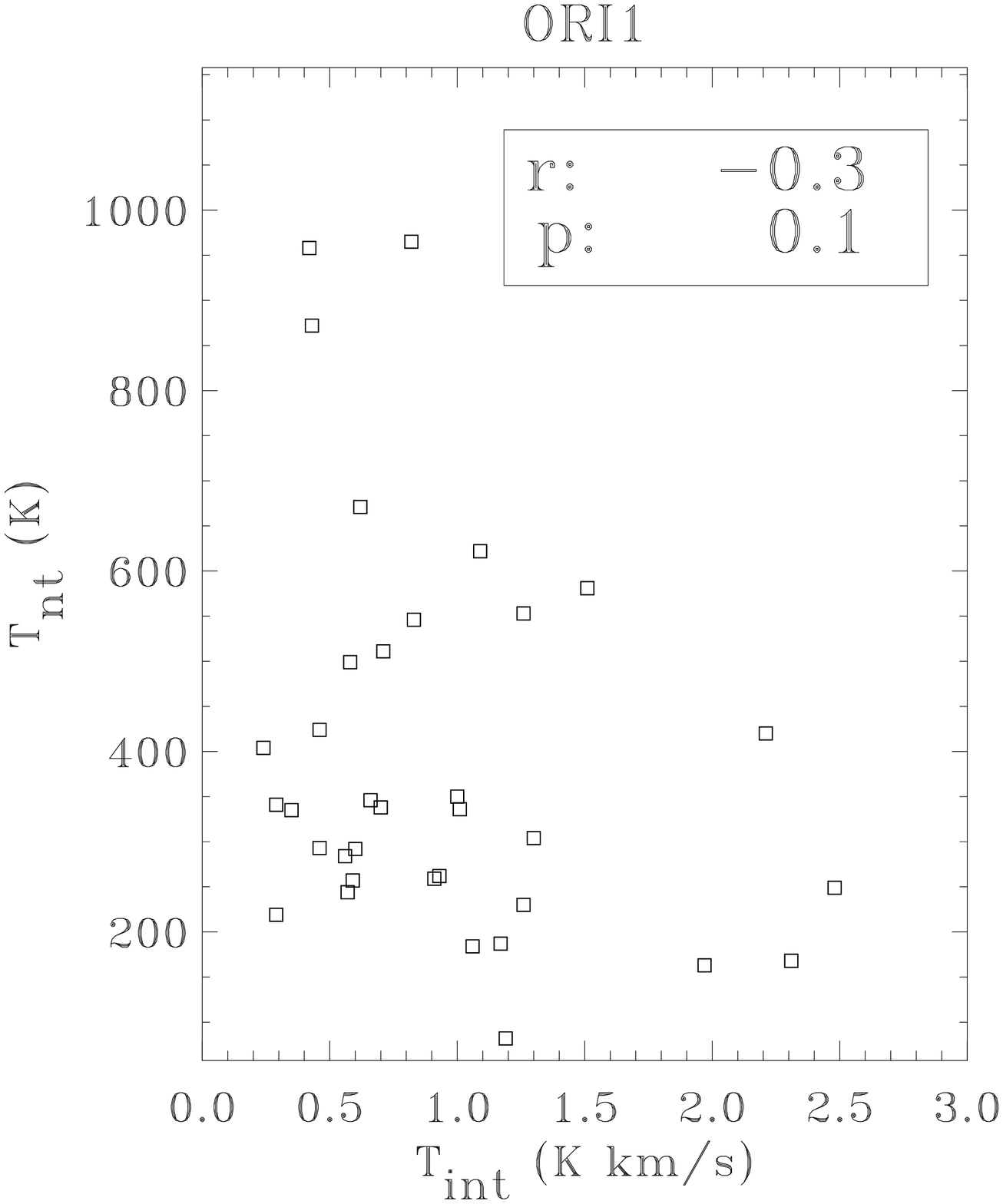}}
     	\subfigure{\includegraphics[width=8cm]{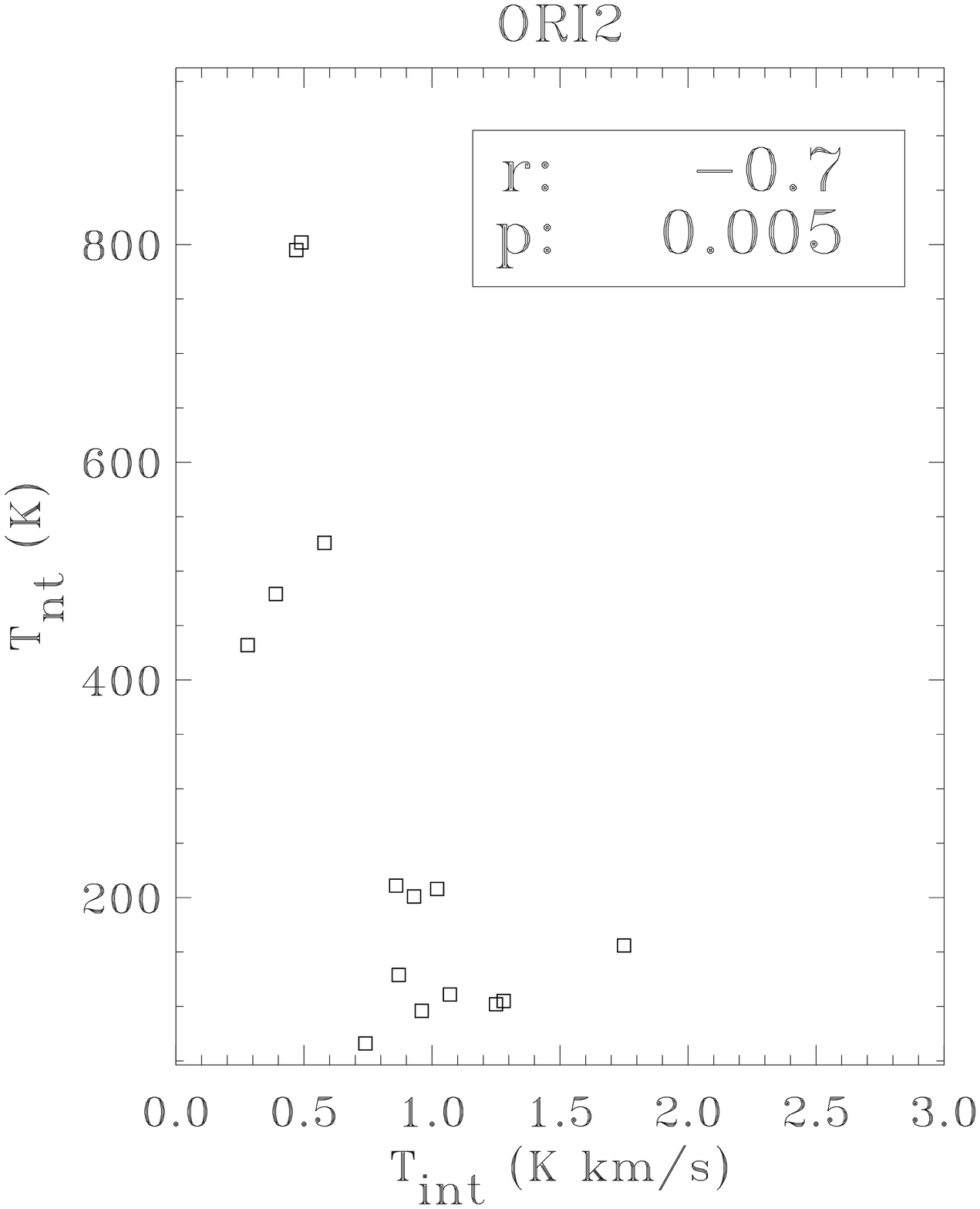}}\\
 	\subfigure{\includegraphics[width=8cm]{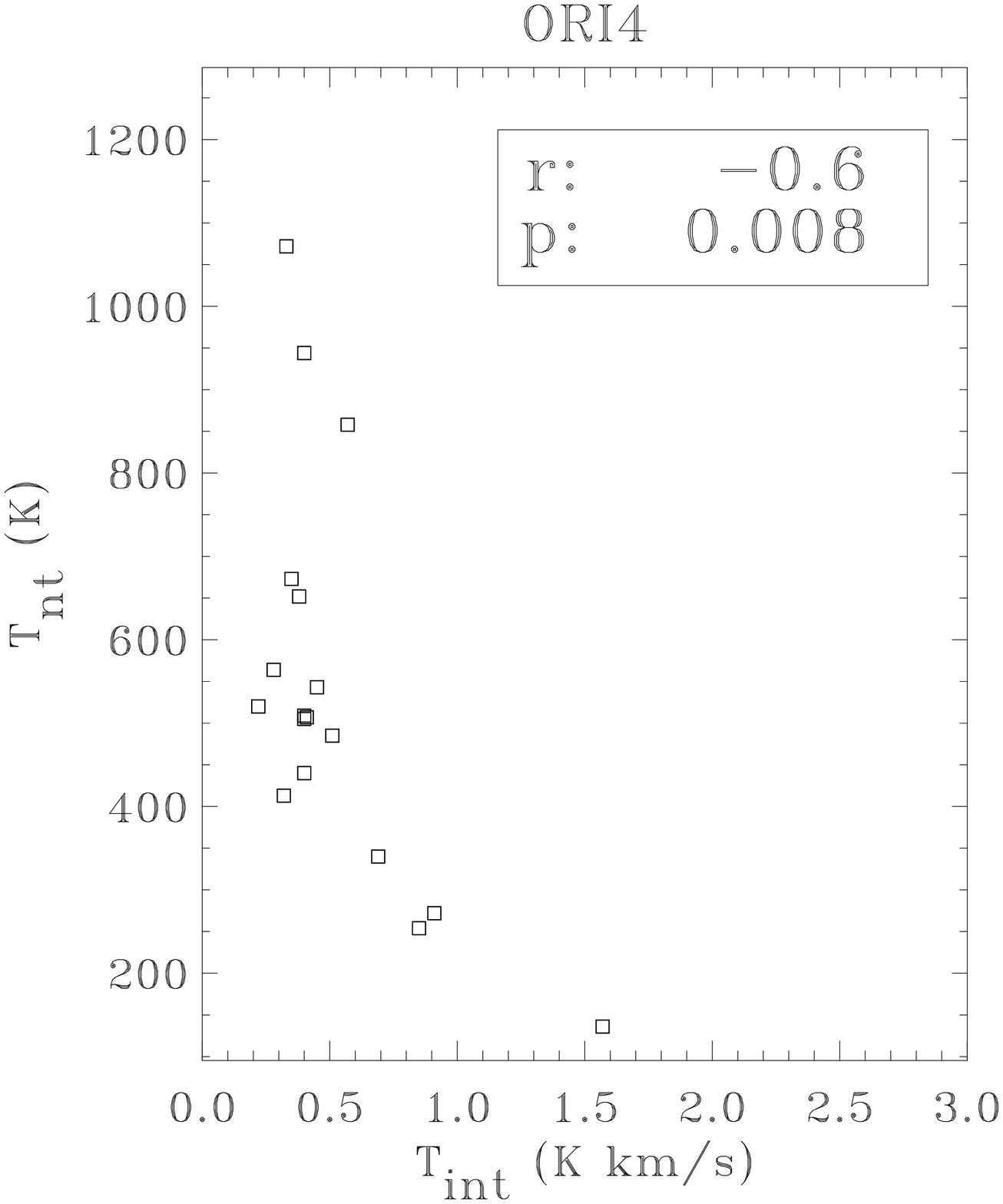}}
     	\subfigure{\includegraphics[width=8cm]{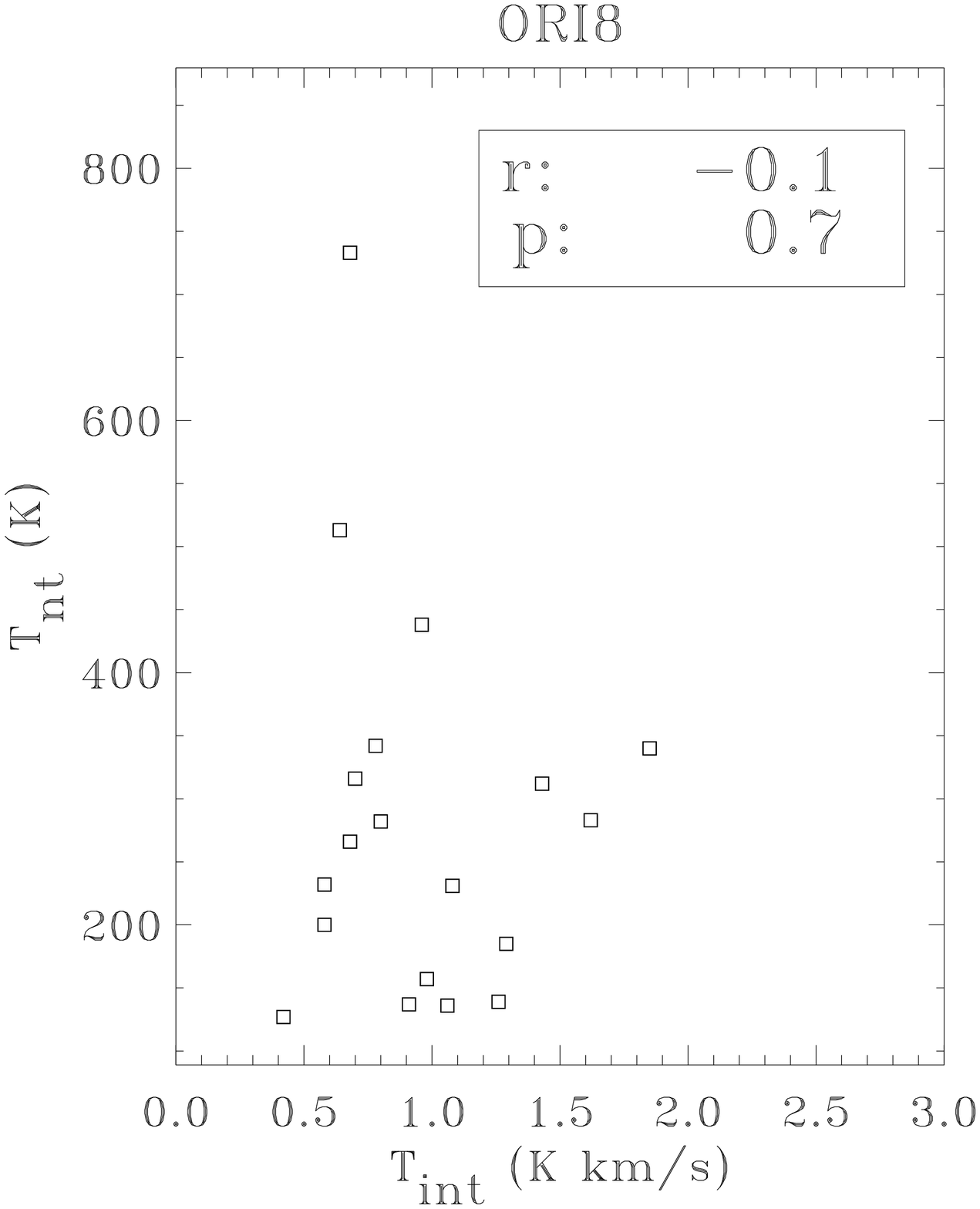}} 
    	\caption{Correlations between integrated intensities of \ammonia\ (1,1)
		line and the turbulence measure $T_{nt}$. 
The Pearson correlation coefficient 
		$r$ and the significance of the null hypothesis $p$ 
		(no correlation) are given in each small box. The uncertainty
in  $T_{nt}$ is about 5 K (see text).} 
    	\label{fig:tur}
\end{figure}

	With the knowledge of kinetic temperature, the turbulence
in these cores
can be quantified in terms of the nonthermal linewidth
\begin{equation}
\label{vnt}
%
%
\Delta V_{nt} = \sqrt{\Delta V^2 - \frac{8 \ln(2) k T_k}{\mu  m_H} } \lc
\end{equation}
where $\Delta V$ is the observed FWHM of the line, $\mu$ is the molecular weight, 
$m_H$ is the hydrogen mass and $k$ is Boltzmann's constant.
To compare it more directly with the thermal motion,
we define
\begin{equation}
T_{nt}= \frac{\mu  m_H}{8 \ln(2) k }\Delta V^2 - T_k \lp
\end{equation}
$T_{nt}$ measures $\Delta V_{nt}$ in Kelvins.
The correlation between the integrated intensities
and the turbulence is presented in Figure~\ref{fig:tur}.
The combined effect of uncertainties in $T_k$ and
 $\Delta V$  amounts to about a 5 K 1$\sigma$
uncertainty in $T_{nt}$.

	The values of $T_{nt}$ range from 100 to 1000 K.
Compared to 10 to 20 K for $T_K$,
there is no doubt that these clouds are dominated by turbulence
throughout. This is in contrast with some dark cloud cores,
which approach being thermalized toward the center \citep{goodman98}.
Three (ORI1, ORI2, ORI4) out of four sources exhibit
an anticorrelation between turbulence and intensity. For ORI8, there is
no apparent correlation. This is similar
to the relations we find for $T_k$. Again, we may not have enough
data on the outer portions, where the emission is weaker.

In this section, we have presented evidence that these massive cores in Orion  
are generally colder inside. They are supersonic, with turbulence
diminishing toward the center of three of the four cores.

\section{Thermal Balance of Externally Heated Cores \label{therm}}

The massive quiescent cores of this study are warmer than molecular regions
away from regions of star--formation and shielded from the interstellar radiation
field (ISRF), which have characteristic temperatures $\simeq$ 10 K.  
Thus, there must be additional heating sources present.
The kinetic temperatures we are considering here come from observations of
\ammonia, which as seen in previous sections, is well--confined to limited
regions within the much larger L1640 and L1641 portions of the Orion molecular cloud.
As such, we are measuring the temperature in regions with considerable
extinction, which are plausibly quite different than the outer parts
of the general molecular material.  
The general picture of the Orion molecular cloud/HII region 
is that the ``front'' of the cloud,
the side facing the Earth, is heated by the radiation from the Trapezium
star cluster, located yet closer to us.  
This is the basis for the model of \citet{stacey93} for the distribution
of ionizing radiation across the surface of the cloud, which is
responsible for the very extended [CII] emission.
The FUV intensity at the interface between the molecular cloud
and the Trapezium is estimated by these authors to be a factor 5$\times$10$^4$
greater than that of the nominal ISRF.

In addition, there will be some heating by the embedded sources. 
For example, \citet{strom93} found $\simeq$ 1500 solar-type PMS stars
distributed throughout the L1641 molecular cloud.  
The region they studied extends over a much larger region than that
defined by the present sample of massive cores. This suggests that
 the heating from embedded PMS stars
as well as  from outflows \citep{morgan91}, while locally possibly significant,
is certainly very inhomogeneous, and in aggregate is not very important.
Thus we will analyze the thermal structure of the cores considering
only the diffuse heating from the Trapezium stars.

Following the treatment of \citet{stacey93}, we take the Trapezium stars
to be located $d_o$ = 0.39 pc in front of the surface of the molecular cloud,
which is assumed to be a plane.  
Then, the radiation field normalized to the standard ISRF at a
 projected distance $d_p$ (in pc parallel to the cloud surface)
from the Trapezium to the point of interest is given by

\begin{equation}
\chi = \frac{8.6\times10^{3}}{(d_o^2 + d_p^2)^{3/2}} ~.
\end{equation}
For ORI1, $d_p$ = 3.2 pc and $\chi$ = 260, for ORI2, $d_p$ = 4.8 pc 
and $\chi$ = 77, for ORI4, $d_p$ = 7.1 pc and $\chi$ = 24, and for
ORI8, $d_p$ = 9.1 pc and $\chi$ = 11. 
These are upper limits, due to 1) the intervening absorption of the
radiation between the Trapezium and the molecular cloud, and 2) the
absorption in the molecular cloud itself.  
Both of these are quite uncertain, particularly as the location of the
cores relative to the surface of the molecular cloud is not known.
We thus consider cores with a range of values of $\chi$
\footnote{
The ``standard'' ISRF of \citet{math83} (MMP) has been reexamined by \cite{black94}
who finds a modest factor greater flux at infrared -- far infrared wavelengths.
The difference in the dust temperature that would result is small
 for $\chi$ = 1. In models of spherical dust clouds by
\cite{evans01}, the corresponding difference is 2.5 K at the cloud 
center, and even less at the
outside of the cloud.
A similar calculation by \cite{zucconi01} gives a difference of about 1 K.
The modeling here is necessarily relatively crude given the uncertainties
in geometry and internal extinction, so that using the MMP radiation
field and multiplicative factor $\chi$ is a reasonable procedure.
}

We have used the radiative transfer code DUSTY \citep{ivezic99}
to calculate the dust temperature within a spherical core having 
specified value of $\chi$ on its periphery.
The form of the external ISRF is given by \citet{math83} for a 
Galactrocentric distance of 10 kpc.
The dust density is uniform throughout the core, and the dust properties
are those of \citet{draine84} with relative number densities
0.53 silicate and 0.47 graphite grains.  
The grain size distribution is given by a power law of index q = 3.5
as discussed by \citet{math77}, with a minimum grain radius $a_{min}$
= 0.005 \mic, and maximum grain radius $a_{max}$ = 0.25 \mic.

\begin{figure}[htp]
      	\centering 
      	\subfigure{\includegraphics[width=.60\textwidth, angle=0]{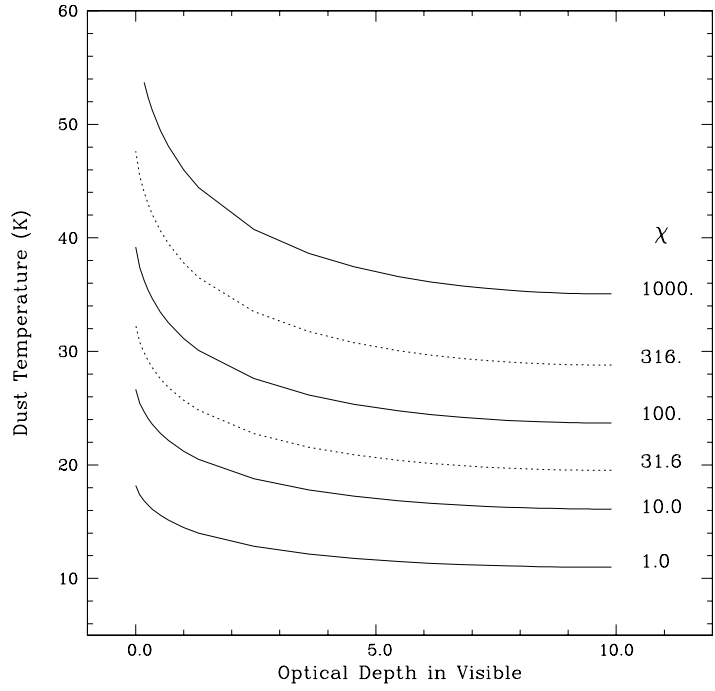}}\\
      	\subfigure{\includegraphics[width=.60\textwidth, angle =0]{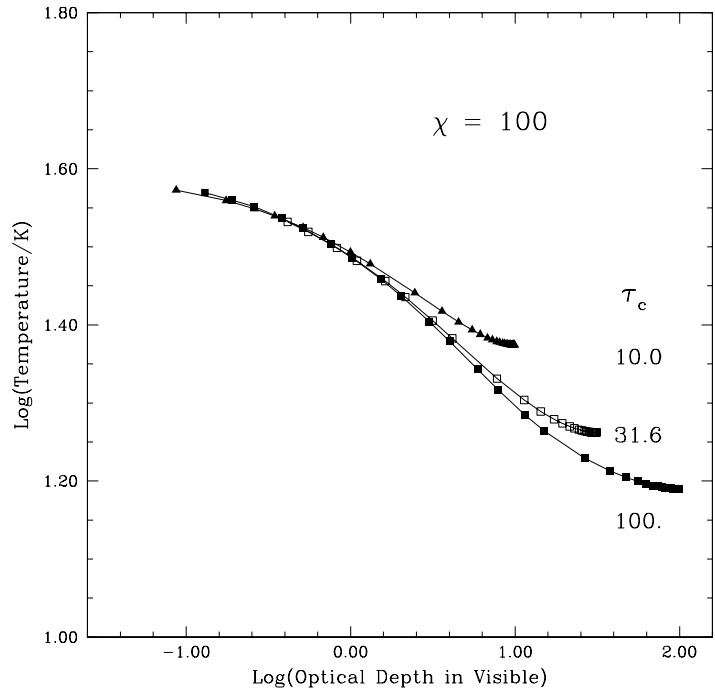}} 

     	\caption{Upper panel:dust temperatures for clouds under different external
     	 	radiation fields.  $\chi$ is the factor by which the standard ISRF is
     	 	multiplied.
     		Lower panel: dust temperatures for clouds with total 
     		center--to--edge optical depths equal to 10, 31.6, and 100.  }
	\label{fig:td-tau-chi}
\end{figure}

The code carries out calculations on a grid of optical depth at
reference wavelength $\lambda_0$ specified to be 0.55 \mic.  
In the upper panel of Figure~\ref{fig:td-tau-chi}, 
we show the dust temperature as a function of the optical
depth in the visible $\tau_{v}$($\lambda_0$).
These calculations are all for a cloud having $\tau_v$
from edge to center equal to 10 (this is somewhat less than
for the massive cloud cores of interest here, but the behavior 
is the same).
We see that the dust temperature is enhanced at the edge of 
the core, but drops significantly when the optical
depth to the edge is a few, and reaches an asymptotic value
for optical depths greater than this.  

\begin{figure}[htp]
      	\centering 
      	\includegraphics[width=.90\textwidth, angle=0]{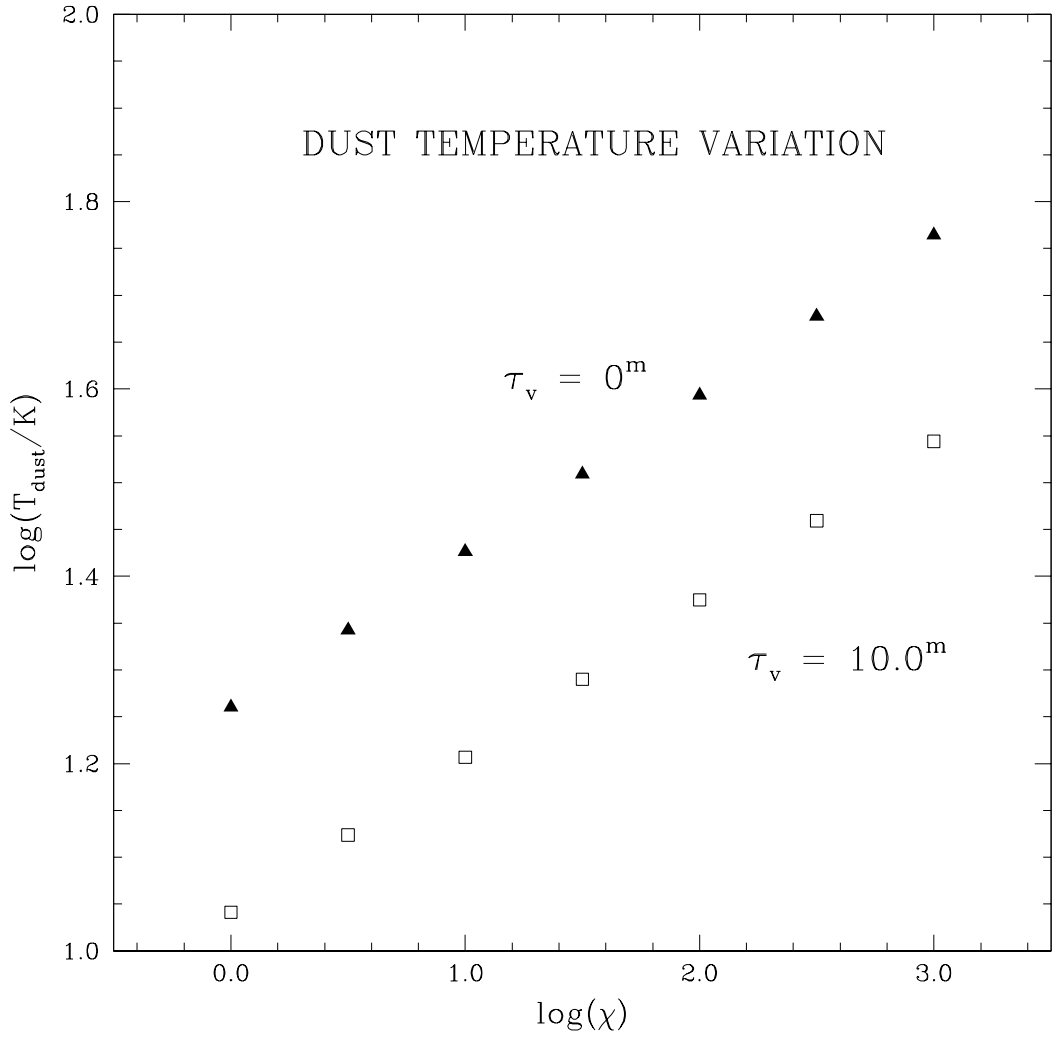} 
     	\caption{Dust temperatures at the cloud surface and at visual optical 
     		depth equal to 10. The power--law dependence of $T_{dust}$ on $\chi$, $T_{dust} \propto \chi^{1/6}$, is expected from the dust model used
     		here (see discussion in the text).}
	\label{fig:td-chi}
\end{figure}

The dust temperature exhibits a very clear variation as a function
of ISRF intensity, both at the core edge, and at its center.  
In Figure~\ref{fig:td-chi}, we show the variation in dust temperature as a function
of $\chi$, for $\tau_v$ = 0 and 10.  
In both cases, the slope of the log--log straight line indicates
that $T_{dust} \propto \chi^{1/6}$.
This is consistent with simplest dust thermal balance consideration,
with dust cooling varying as $T_{dust}^6$ and dust heating being
proportional to $\chi$.
This cooling law is what is expected if the dust emissivity throughout
the wavelength range of interest varies as $\nu^2$ \citep{goldsmith01}.
While the dust emissivity in the Draine \& Lee (1984) model does
not obey this law exactly throughout the millimeter--to--infrared
range, it is sufficiently close at these modest temperatures that the
cooling variation is very accurately represented by a $T_{dust}^6$
power law.

It is reasonable that the dust heating should be proportional to
$\chi$ for small $\tau_v$, but it is somewhat surprising that
this continues when the optical depth
between the external heating source and the dust is much greater than
unity.
This can be understood, however, as a consequence of the heating of
the dust by radiation emitted by other dust grains in the cloud.
The total heating of a grain at the cloud center still varies in
proportion to $\chi$ even though the wavelengths at which the heating
occurs are much longer than for direct heating by the ISRF at low
optical depth.

The exact values of the dust heating and hence the dust temperature
depend on the total optical depth of the core as well
as the optical depth to the core edge.
This is seen in the lower panel of Figure~\ref{fig:td-tau-chi}, 
which shows the variation in dust temperature
for $\chi$ = 100, as a function of $\tau_v$, for three clouds having
different values of central optical depth $\tau_c$.
As expected, the dust temperature is independent of $\tau_c$ for small
values of $\tau_v$, since the heating is dominated by direct input from
the ISRF.
As the dust heating makes a transition to being dominated by reradiation
from other dust grains in the core, we see a systematic effect, which
is that the dust temperature decreases as the central optical depth $\tau_c$
increases.  
Thus, a more opaque cloud is characterized by cooler dust in its interior.
This can be seen to be a result of balance between total cooling of the
cloud, which (for its interior) is dominated by emission at
relatively long wavelengths, which will not be highly optically thick.  
If we consider two clouds of the same size, they intercept the same total
energy from the ISRF, so that total heating is the same.  
The cooling will, however, be greater for the cloud of higher opacity,
since it will radiate with greater or equal efficiency at all wavelengths,
and thus the dust temperature within it will be lower.

Thus, while we can use simple scaling rules to determine dust temperature
variation as a function of $\chi$, the exact value of $T_{dust}$ 
depends on position in the core and the total optical depth of the core. 
From column 7 of Table 1, we see that the \c18o column densities for
the four cores studied in detail (ORI1, ORI2, ORI4, and ORI8) are in 
the range $\simeq$ 5 -- 13 $\times$10$^{15}$ \cm2,
which for a canonical fractional abundance X(\c18o) = 10$^{-7}$ and
with $\tau_v$ = 10$^{-21}$N(H$_2$) implies edge--to--center visual
optical depths ($\tau_c$) between 25 and 65.
Using  $\tau_c$ = 31.6 as a representative value, we find that
the dust temperature, T$_d$, at the edge and center, respectively, are 26 K and 12 K 
for $\chi$ = 10, 38 K and 29 K for $\chi$ = 100, and 56 K and 28 K for $\chi$ = 1000.


The gas temperature, T$_g$, is affected by a variety of processes.  
In the well--shielded centers of these cores, the primary effects are
cosmic ray heating, molecular line cooling and collisional heating 
due to dust--gas collisions.
Using the parameters from \cite{goldsmith01} with no depletion,
for T$_d$ = 20 K, we find that T$_g$ = 13.5 K for n(H$_2$) = 10$^4$ \cc\
and 18 K for n(H$_2$) = 10$^5$ \cc.
Thus, given the molecular hydrogen densities for these cores, $\simeq$ 4 -- 20
$\times$10$^4$ \cc, the central gas temperature, which is the kinetic
temperature determined from the \ammonia\ observations, should be a few K lower
than the dust temperature.
The situation at the edges of the cores is more complicated, as the fractional
abundance of molecular coolants will drop, while the abundances of atomic and
ionic coolants will increase.  
The density may also plausibly be lower, although the degree of central
condensation of these cores has not been determined.

Comparison of observational results of core ``central'' temperatures from our \ammonia~ data
with the expectations from external dust heating shows reasonable 
agreement.
For ORI1, ORI2, and ORI4, the kinetic temperature is 2 -- 4 K below that expected
for the dust, which is consistent with the thermal balance models and core
densities (Table 1).
For ORI8, the kinetic temperature is a few K higher than the dust temperature
predicted by the model.  
The data for this source are the least extensive of the mapped cores, but the
central portion does appear to be well defined (see Figure 2).
It is possible that some other modest heating source is raising the
temperature of this core.

The ``external'' core temperatures found by \cite{wilson99} are in surprisingly
good agreement with the dust temperature predicted from heating by the Trapezium.
It is plausible that the $^{12}$CO samples the temperature in a relatively
thin shell at the periphery of the core.
This can be seen from evaluating the standard expression for the optical
depth, which for $T_k$ = 20 K and \nh = $10^3$ \cc, yields for $^{12}$CO
with a Gaussian line profile having FWHM $\delta v$ \kms
\begin{equation}
\tau_{1,0}(^{12}\text{CO}) = 5.5\times10^{-17} \frac{ N(^{12}\text{CO})/\text{\cm2}}{\delta v /\text{\kms} } \lp
\end{equation}
For a nominal fractional abundance, N($^{12}$CO) = $10^{-4}$N(\h2) = 
$10^{17}\tau_v$.
Substituting this in the preceding equation, we see that for a FWHM line
width equal to 2 \kms,
\begin{equation}
\tau_{1,0}(^{12}\text{CO}) \simeq 3 \tau_v\lp
\end{equation}
The optical
depth of the J = 2--1 transition under these same conditions 
is a factor of 2.6 greater than that of the 1--0 transition, and
$\tau_{3,2}$ = 3$\tau_{1,0}$.
Thus, if a fractional abundance X($^{12}$CO) = 10$^{-4}$ prevails right to the edge of the
cloud, the $^{12}$CO rotational transitions become optically thick in a small 
fraction of the thickness required to have a dust optical depth of
unity in the visible.  
However, the fractional abundance of $^{12}$CO drops in the outer portion
of the core due to photodestruction, and despite its own self--shielding,
its fractional abundance becomes significant only when $\tau_v$ approaches
unity.
In any case, it appears secure that the J = 3--2 transition of $^{12}$CO
as observed by \cite{wilson99} will probe the gas temperature in the
outer layer of the core.
The density there may be inadequate for perfect collisional gas--dust coupling, but
other processes, including photoelectric heating, will help raise the
gas temperature. 
Thus, although the close agreement between the dust temperature predicted
by the external heating models, and the temperature derived from J = 3--2
$^{12}$CO observations may be somewhat fortuitous, it appears that 
we can use the temperature from $^{12}$CO as a probe of the outer
region of the core
\footnote{
The issue of externally heated material in GMCs, particularly Orion, is
not new, but is not always recognized in analyses of multi--transition 
studies and maps. \citet{castets90} inferred that the molecular gas in
Orion is hotter on the outside than in its interior, based on arguments
from densities obtained from the two lowest transitions of $^{13}$CO, in
a manner similar to that found for dark clouds by \citet{young82}. 
\citet{tauber90} found that their observations of a ratio $^{12}$CO J = 3--2
to J = 1--0 $\simeq$ 1.4 could be reproduced by hot--edged clumps.
\citet{gierens92} developed a detailed model of CO emission
from a clump with photoelectric heating by a strong external radiation field,
and find that this can satisfactorily reproduce the observations of
\citet{castets90}.  
The $^{12}$CO emission reflects the $\simeq$ 30 K temperature
at the inner edge of the PDR bounding the molecular portion of the clump.  
This is considerably warmer than the interior
of the clump, which contributes significantly to the $^{13}$CO emission.
}.

Overall, the temperatures known from CO and \ammonia\ observations
can be well explained by the dust being heated by an enhanced external UV field.

\section{Summary and Discussion \label{diss}}

In this paper we have presented the first portion of the results of a study of
dense cores in the nearby Orion Giant Molecular Cloud.  
The cores selected are apparently free of signposts of star formation (embedded
sources or outflows) and which are relatively distant from the perturbing effects
of the formation of massive stars.
We have focused on the temperature structure of these cores, primarily as
traced by \ammonia.
We have also utilized other parameters obtained from mapping in \c18o,
which will be presented in detail elsewhere. 
In particular, the mean density of the cores is $\simeq$ 4 -- 20 $\times$10$^4$ \cc.
This is significantly higher than the mean density of the molecular cloud at
these large distances ($\simeq$ 20\arcmin\ to 90\arcmin; $\simeq$ 4 to 14 pc) from its center
\footnote{
Multitransition determinations of densities have yielded comparable or higher values
over regions extending $\leq$ 10\arcmin~ from the cloud center \citep{bergin96}.
However, comparison of these densities with those obtained from virial arguments
\citep{goldsmith99} reinforces the picture that there is a filling factor of 
high density material which is relatively large near the center of the GMC, but
drops significantly at larger distances \citep{mundy86}.
At the distances from the Trapezium of the cores studied here, the average cloud 
density is n(H$_2$) $\simeq$ few$\times$10$^3$ \cc.
}.
The density contrast between these cores and the surrounding material is at least
an order of magnitude.

We have discussed in some detail the issue of the temperature structure of
these cores.  
They appear as significantly cooler than the surrounding material.
This is seen in two different ways.
First, the kinetic temperature derived from the \ammonia\ (1,1) and (2,2) lines
drops from the edges to the centers of the cores.  
Their central temperatures are between 14 and 19 K, for the four cores studied
in greatest detail.  
The core edges are  between 4 and 8 K warmer than the centers.
The ability to probe the core temperature structure solely with \ammonia\ is
limited by the dramatic decrease in its emission as a function of increasing
distance from the core center.
Second, we can get a better picture of the overall thermal structure by including 
the kinetic temperature derived from $^{12}$CO, which samples an outer ``onion skin'' 
of the core.
Using J = 3 --2 data from the literature, we find temperatures between 18 and 33 K.
This is consistent with model calculations of dust heating by a single source,
the Trapezium, the radiative transfer within the dust of the core, and dust--gas
coupling via collisions.

These cores, although colder than their surroundings, are still
supersonic. They are more quiescent than their surroundings with
turbulence level dropping toward the center, as indicated by the reduced nonthermal 
width of the \ammonia\ emission lines.
Since the kinetic temperature has been derived independent of linewidth,
this result has been established with confidence.
This first part of a study of dense cores in HMSF molecular clouds has established some
important parameters of these potential sites of star formation.
The data and analysis presented in a subsequent paper will address the density,
structure, and energetics of these regions, confirming them as plausible, relatively
isolated sites of future formation of high mass stars.

\acknowledgments
We thank P. Schilke and D. Muders for their kind help with the 
telescope operation and data calibration of the Effelsberg 100--meter telescope.
Our \ammonia\ observations were the last K--band undertaking at the 140--foot
telescope, and they were only made possible with assistance from D. Balser.
The National Astronomy and
Ionosphere Center is operated by Cornell University under a Cooperative
Agreement with the National Science Foundation.  The Five College Radio
Astronomy Observatory is supported by NSF grant AST97-25951.


\begin{deluxetable}{lcccccccccccccc}
\rotate
\tabletypesize{\scriptsize}
\tablewidth{0pt}  
\setlength{\tabcolsep}{0.04in}  
\tablecolumns{14}  
\tablecaption{Core Parameters}
\tablehead{  
\colhead{Source} 	&\colhead{RA(1950)\tablenotemark{a}} 
			&\colhead{DEC (1950)\tablenotemark{a}} 
			&\colhead{ V$_{LSR}$\tablenotemark{a}}
			&\colhead{$\Delta$RA\tablenotemark{b}} 
			&\colhead{$\Delta$DEC\tablenotemark{b}}  
			&\colhead{N(\c18o)} 
			&\colhead{N(\ammonia)\tablenotemark{c}} 
			&\colhead{$\Delta$V(\ammonia)\tablenotemark{c}} 
			&\colhead{T(\ammonia)\tablenotemark{d}} 
			&\colhead{T(CO) \tablenotemark{e}}
			&\colhead{$d_p$\tablenotemark{f}}  
			&\colhead{Radius\tablenotemark{g}} 
			&\colhead{Mass} 
			&\colhead{n(H$_2$)\tablenotemark{h}} \\
			&\colhead{ h m s}
			&\colhead{d m s }   
			&\colhead{km s$^{-1}$}
			&\colhead{arcmin}
			&\colhead{arcmin}
			&\colhead{10$^{15}$ \cm2}
			&\colhead{10$^{14}$ \cm2}
			&\colhead{km s$^{-1}$ }
			&\colhead{K}
			&\colhead{K}
			&\colhead{arcmin}
			&\colhead{arcmin}  
			&\colhead{\Ms} 
			&\colhead{$10^4$ \cc}} 
\startdata
ORI1&5 32 49.7  &-5 02 23  	&10.6 	&0.0 & 0.0 &13$\pm2$\tablenotemark{i} & 3.3$\pm1.3$\tablenotemark{i}&0.85$\pm0.07$\tablenotemark{i} & 19$\pm1$\tablenotemark{i} &33& 23&1.4 &260 & 20\\
ORI2&5 32 36.3	&-5 59 43 	&8.4  	&2.0 & 1.5 &6.5& 2.7 &0.68  & 15& 20&34	&2.6 &360 &4.5\\
ORI3&5 32 44.3	&-6 01 43 	&8.2  	&NA &NA  &NA &NA&NA&NA&25&36	 &NA &NA &NA\\
ORI4&5 33 37.9	&-6 14 23 	&9.1  	&4.6 &-0.7  & 5.3& 4.5 & 0.63&15& 23& 51&2.3 & 250& 4.8\\
ORI5&5 34 10.2	&-6 15 03 	&8.2  	&-3.9 &0.0  & 3.1& 0.15& 0.88&14& 22& 54&\nodata &\nodata &\nodata\\
ORI6&5 34 02.0	&-6 17 03 	&8.2  	&NA &NA &NA &NA & NA& NA&18&55 &NA &NA &NA\\
ORI7&5 33 27.3	&-6 28 23 	&8.8  	&2.8 &2.2  &4.5 & 0.30& 0.69&14 & 28&64 &\nodata &\nodata &\nodata\\
ORI8&5 34 18.2	&-6 26 23 	&8.2  	&1.5 &-5.6  &5.6& 4.4& 0.73&15 &18& 65&3.1 &490 &3.6\\
ORI9&5 33 38.0	&-6 31 43 	&8.4  	&-1.5 &-2.2 &3.5 &\nodata &\nodata & \nodata& 35&68& \nodata & \nodata &\nodata\\
ORI10&5 33 54.1	&-6 35 03 	&6.8  	&-0.7 &2.2  & 4.1 &\nodata &\nodata & \nodata& 12& 72&1.9 & 106&3.5\\
ORI11&5 33 51.3	&-6 41 43 	&7.0 	&0.7&-1.5  &2.9  &\nodata &\nodata & \nodata & 18&78 & 1.2& 60&8.4\\
ORI12&5 33 48.8	&-6 50 23 	&8.4  	&0.7 &2.9  & 8.8 &\nodata &\nodata & \nodata& 14& 86&\nodata &\nodata &\nodata\\
ORI13&5 36 03.3	&-7 18 23 	&6.1  	&-1.5& 0.0 & 2.3 &\nodata &\nodata & \nodata& 15&123 &\nodata & \nodata&\nodata\\
ORI14&5 36 24.0	&-7 26 23	&5.4  	&2.2 &2.9  & 3.3 &\nodata &\nodata & \nodata& 10&132 &\nodata &\nodata &\nodata\\
ORI15&5 36 10.6	&-7 28 23	&5.6  	&0   & -2.2& 3.2 &\nodata &\nodata & \nodata&10&133 &\nodata&\nodata &\nodata\\  
\enddata

\tablenotetext{a}{Data are taken from the Orion CS survey \citep{tatematsu93}.}
\tablenotetext{b}{The Offsets given in columns 2 and 3 are measured from the peak of 
	the \c18o integrated intensity to that of the CS. For ORI3 and ORI6, their 
\c18o\ peaks (denoted as ``NA'') are located at the same places as Ori2 and Ori4, respectively.
This notation applies to other columns as well.}
\tablenotetext{c}{Values refer to the positions of strongest ammonia emission.
 For ORI5 and ORI7, the values are
given for the (0,0) positions as given in columns 2 and 3 since \ammonia\ map is not available
for these two sources. For all other source (``...''), the (2,2) line is not detected 
at a noise level of 0.03 K RMS. 
}
\tablenotetext{d}{Kinetic temperature obtained from our ammonia observations. The values
	given refer to the positions of strongest ammonia emission, except for ORI3 and ORI6.}
\tablenotetext{e}{Kinetic temperature based on CO 3-2 spectra \citep{wilson99}.}
\tablenotetext{f}{The projected distance from the source to the Trapezium cluster.}
\tablenotetext{g}{The square root of the product of semi--major and semi--minor axes. This size parameter
is only given for cores whose \c18o\ half intensity contours are enclosed by our 6\arcmin$\times$6\arcmin\ 
maps. The same is true for the last two columns of the table.}
\tablenotetext{h}{Mean molecular hydrogen density obtained from \c18o column density.}
\tablenotetext{i}{The one $\sigma$ statistical uncertainty resulting from a Gaussian fit to the spectra. 
The percentage is representative for this column. The uncertainty for the  kinetic temperature is
discussed in detail in section~\ref{stat}.
}
\label{tab:source}
\end{deluxetable}

\end{document}